# Interaction-Driven Giant Electrostatic Modulation of Ion Permeation in Atomically Small Capillaries

Dhal Biswabhusan[1], Yechan Noh[2], Sanat Nalini Paltasingh[3], Chandrakar Naman[1], Siva Sankar Nemala[4], Rathi Aparna[1], Kaushik Suvigya[1], Andrea Capasso[4], Saroj Kumar Nayak[3], Li-Hsien Yeh[5,6], Kalon Gopinadhan*[1,7]

[1]Department of Physics, Indian Institute of Technology Gandhinagar, Gujarat 382355, India

[2]Department of Physics, University of Colorado Boulder, Boulder, CO 80309, USA

[3]School of Basic Sciences, Indian Institute of Technology Bhubaneswar, Khorda, Odisha-752050, India

[4]International Iberian Nanotechnology Laboratory, Braga 4715-330, Portugal

[5]Department of Chemical Engineering, National Taiwan University of Science and Technology, Taipei 10607, Taiwan

[6]Advanced Manufacturing Research Center, National Taiwan University of Science and Technology, Taipei 10607, Taiwan

[7]Department of Materials Engineering, Indian Institute of Technology Gandhinagar, Gujarat 382355, India

(*Corresponding author Email: gopinadhan.kalon@iitgn.ac.in)

**Manipulating the electrostatic double layer and tuning the conductance in nanofluidic systems at salt concentrations of 100 mM or higher has been a persistent challenge. The primary reasons are (i) the short electrostatic proximity length, ∼3-10 Å, and (ii) difficulties in fabricating atomically small capillaries. Here, we successfully fabricate in-plane vermiculite laminates with transport heights of ∼3-5 Å, which exhibit a cation selectivity close to 1 even at a 1000 mM concentration, suggesting an overlapping electrostatic double layer. For gate voltages from -2 V to +1 V, the $K^+$-intercalated vermiculite shows a remarkable conductivity modulation exceeding 1400% at a 1000 mM KCl concentration. The gated ON/OFF ratio is mostly unaffected by the ion concentration (10-1000 mM), which confirms that the electrostatic double layer overlaps with the collective ion movement within the channel with reduced activation energy. In contrast, vermiculite laminates intercalated with $Ca^{2+}$ and $Al^{3+}$ ions display reduced conductance with increasing negative gate voltage, highlighting the importance of ion-specific gating effects under Å-scale confinement. Our findings contribute to a deeper understanding of electrostatic phenomena occurring in highly confined fluidic channels, opening the way to the exploration of the vast library of two-dimensional materials.**

**Introduction**

Selective and tunable ion transport is crucial in advanced material applications, including 'lab-on-a-chip' devices[1,2], energy harvesters[3,4], desalination membranes[5,6] and neuromorphic devices[7,8]. Among the many external stimuli affecting ionic transport, electrostatic gating appears to be highly reversible. Inspired by the remarkable efficiency of biological channels that rely on gated ion transport to activate certain functions[9], researchers have explored the effects of electrostatic gating in several confined systems[10]. Although few experimental groups have achieved reasonable gate modulation at dilute ion concentrations, little progress has been made at physiologically and practically relevant concentrations[11,12,13] (i.e., 100 mM or higher). To be effective at high ion concentrations, the channels need to overcome electrostatic screening, which essentially restricts the channel size available for ion transport. The Debye length estimation at a 1000 mM KCl concentration suggests that the channel size should be smaller than 6 Å. These challenging size requirements explain the absence of works addressing high concentrations. Nevertheless, gating studies at dilute concentrations revealed several intriguing effects: dehydration of ions[14], modification of interlayer interactions[15], concentration polarization[16], and ion sieving[17,18]. In the past, several confined systems have been explored, such as silica nanochannels and nanowires[11], unintercalated graphene oxide (GO laminates[12]), MXenes[13], MXene-GO [Ref.[19,20]] and $TiO_2$ nanochannels[21]. However, the majority of these systems have channel sizes > 7 Å. The best gate modulation effect reported thus far is < 1000% at concentrations as small as 10 mM.

At a high salt concentration (1000 mM), significant ion interaction effects are expected if the channel size is comparable to the Debye length, as demonstrated by molecular dynamics (MD) simulations[22,23]. At this length scale, ion-ion and ion-water-ion interactions are likely to occur because of the close proximity of ions, water, and surfaces. Moreover, experiments can detect and study these effects. For example, the knock-on mechanism of ion transport due to Coulomb repulsion could lower the energy barrier and enhance the ionic conductance[24]. Electrostatic gating of the fluidic surface can change the ionic concentration and their transport in highly confined channels. In addition to gating, the valency of the ions can also affect their interactions. In the very few studies, electrostatic gating was also found to alter ion–water interactions[14].

Modulating ion transport at sea salt concentrations of 600 mM or higher is extremely difficult because of the requirement of highly confined fluidic channels, since the short Debye screening length at this concentration makes electrostatic gating ineffective. The focus of this paper is to modulate ion transport at these extreme concentrations, which is rarely discussed. There are many advantages of having high ion concentrations; for example, a high concentration ensures high ionic conductivity and hence a better signal-to-noise ratio. High concentrations may provide better stability against environmental fluctuations.

With this motivation, we designed channels with an overlapping electrostatic double layer (EDL) at high salt concentrations to assess the ion modulation and interaction effects. For this purpose, we fabricated laminates with angstrom-sized fluidic channels using earth-abundant and cost-effective vermiculite clay material. In vermiculites, the interlayer cations can be easily exchanged, thus providing a tunable transport height from ~3 to 5 Å. The atomically small interlayer space ensures that the Debye layer overlaps even at a 1000 mM concentration, enabling the selective and tunable transport of smaller cations such as $K^+$ via electrostatic gating. The insulating membrane itself can serve as a gate dielectric, eliminating the need for an extra layer. The membrane is stable under aqueous conditions[25] and can operate at elevated temperatures[26].

**Results and Discussion**

Vermiculite crystals that occur naturally are composed of layers of magnesium aluminosilicate. Each layer is made up of three sheets: one octahedral sheet occupied by the $Al^{3+}$ ion at the octahedral site can be substituted with $Mg^{2+}$ or $Fe^{2+}$, which is sandwiched between two tetrahedral sheets occupied by the $Si^{4+}$ at the tetrahedral site (Fig. 1a) [Ref.[27,28]]. Additionally, $Al^{3+}$ ions substituted one-fourth of the $Si^{4+}$ ions in tetrahedral sites, causing an enhanced negative charge in the layered structure. These excess charges are balanced by cations such as $Mg^{2+}$ or $Ca^{2+}$ ions, which reside in the space between these layers. In our study, these $Mg^{2+}$ cations were exchanged with other cations via the 2-step heat exchange approach described in the methods section. We intercalated $K^+$, $Ca^{2+}$, and $Al^{3+}$ ions into vermiculite (V) laminates to achieve variable interlayer spacing. To estimate the interlayer spacing ($d$) of vermiculite laminates, we performed X-ray diffraction (XRD) on free-standing Li–V, K–V, Ca–V, and Al–V membranes in both dry and wet states (Supplementary Fig. 1a). We observe no significant difference in the peak positions of both states. Since we performed our ion transport measurements in a wet state, it is appropriate to discuss the XRD results in the wet state. The XRD data revealed a '$d$' spacing of 12.2 Å for K–V. After subtracting the space occupied by the single magnesium aluminosilicate layer, which is 9.6 Å [Ref.[26]], we obtained a transport height of 2.6 Å for ion transport, which is occupied by solvated $K^+$ ions. Similarly, for the Ca–V and Al–V membranes, the transport height is 5.6 Å and 5.1 Å, respectively, occupied by hydrated $Ca^{2+}$ and $Al^{3+}$ ions. In the case of the Ca–V and Al–V membranes, higher-order peaks were observed, indicating complete laminated frameworks. SEM cross-sectional images further confirmed the laminated structure (Supplementary Fig. 2) and helped us estimate the thickness of the membrane (i.e., ∼ 3.5 μm). The cross-sectional energy dispersive (EDS) analysis of the K-intercalated vermiculite membrane revealed the main elements of vermiculite, such as O, Si, Mg, and Al, along with the intercalant K (Fig. 1b). The EDS maps of Ca–V and Al–V similarly show the intrinsic elements of vermiculite along with the intercalant (Supplementary Fig. 3a&b). However, intercalant Al is hard to distinguish from intrinsic Al. The membrane consists of monolayers with a thickness of 14 Å, as estimated from the AFM height profile (Supplementary Fig. 4). The mica-like structure adsorbs 1 or 2 layers of water molecules on its surface and hosts intercalated ions on its surface[29]. This results in a slightly larger thickness of 14 Å instead of 9.6 Å.

Surface charges play a crucial role in controlling nanofluidic transport because the charged solid surface and electrolytes form an electric double layer (EDL) [Ref.[30]], as proposed by Helmholtz in 1853. V-laminates feature a negative surface charge that draws cations to the interlayer space from the electrolytes, creating an EDL, which is confirmed by our zeta potential (Supplementary Fig. 5) and transport measurements. We employed several charged cation intercalants to control these zeta potentials. The zeta potential measurements of the K–V, Li–V, Ca–V, and Al–V solutions indicate that the K–V layers have a larger zeta potential of –50 ± 3 mV than the Ca–V (–19 ± 3 mV) and Al–V (–4 ± 1 mV) layers. We also measured the zeta potential of the membranes; the values were very similar to those of the dispersed solution. The higher zeta potential value of the K–V solution indicates that more $K^+$ is needed to neutralize the surface than $Ca^{2+}$ and $Al^{3+}$. This observation also suggests that in vermiculite, the surface charge can be easily controlled by the appropriate exchange of ions. The vermiculite membrane exhibited mechanical stability, with a tensile strength of ∼ 35 MPa and a fracture strain of ∼ 2% (Supplementary Fig. 6). This demonstrates that vermiculite is flexible and suitable for use in aqueous conditions.

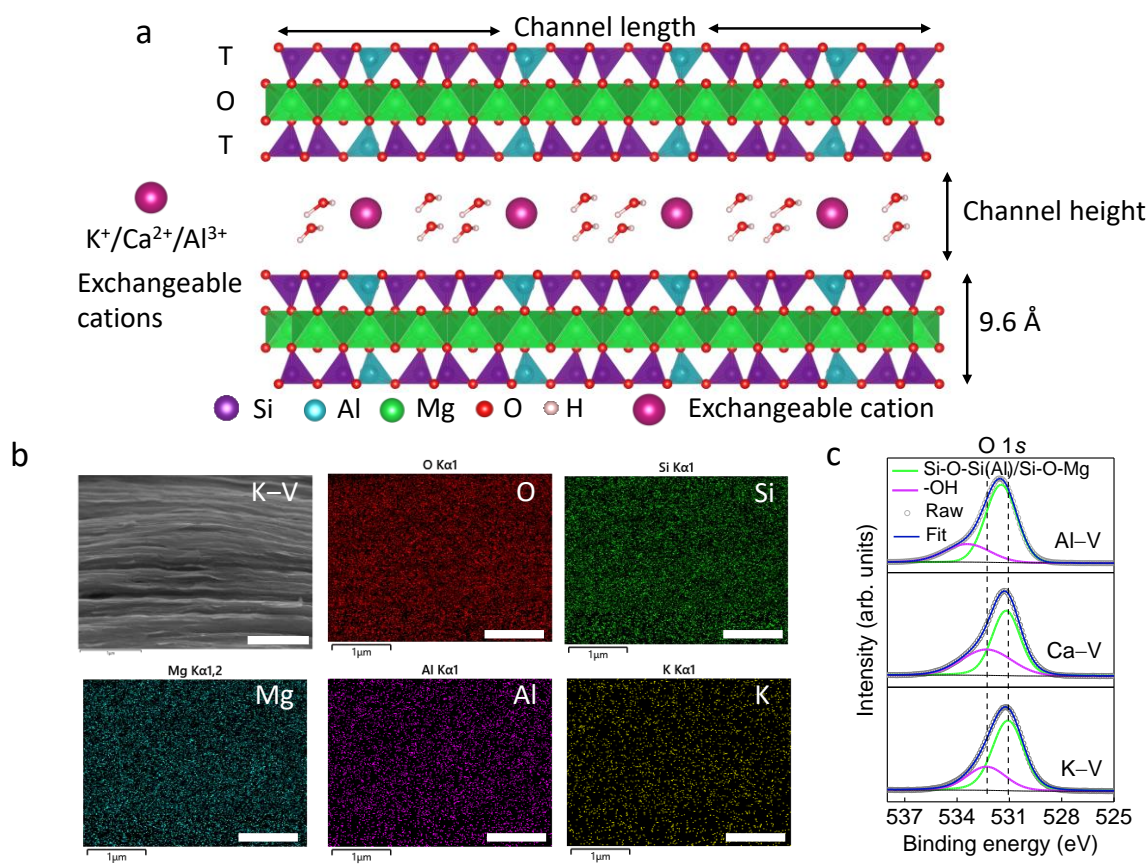

**Fig. 1 | The structure of vermiculite and its composition.** a Schematic of the vermiculite layers. The unit cell consists of tetrahedral and octahedral layers with $Mg^{2+}$ occupying the octahedral sites and $Si^{4+}$ and $Al^{3+}$ the tetrahedral sites with shared oxygen. Exchangeable cations are occupied between the layers to balance the excess negative surface charge and are replaced with $K^+/Ca^{2+}/Al^{3+}$ cations. b Energy dispersive (EDS) analysis of the elements across the layers of the K–V membrane indicates the presence of O, Si, Mg, Al, and the intercalant K. The scale bar indicates 1 μm. c High-resolution X-ray photoelectron O 1*s* spectra (XPS) of the K–V, Ca–V, and Al–V membranes. The grey open circle indicates the raw data of O 1*s*, and the blue line is the fitted data. Two deconvoluted spectra (green line and violet line) were derived from the raw data. The two black dotted straight lines indicate the peak position of -OH (violet) and Si-O-Si(Al)/Si-O-Mg (green) for K-V membranes. A clear shift of peak position towards higher binding energy is observed with increasing cation valence. Source data are provided as a Source Data file.

Furthermore, we carried out a Fourier transform infrared (FTIR) spectroscopy study to confirm the presence of functional groups in the vermiculite. As described in Supplementary Fig. 7, the broad absorption peaks at 3360 cm$^{-1}$ and 1643 cm$^{-1}$ indicate the stretching and bending vibrations of the -OH groups, respectively[31]. The absorbance peak at 970 cm$^{-1}$ corresponds to the asymmetric stretching vibration of Si-O-Si. The two peaks at 735 cm$^{-1}$ and 652 cm$^{-1}$ are related to the Al-O-Al and M-O-Si (M is Mg, Al, and Fe) bonds, respectively[31,32]. We also performed Raman analysis for all the cation-intercalated membranes, which revealed the characteristic peaks of Mg-O/Al-O and Si-O (Supplementary Fig. 8) [Ref.[33]].

We performed X-ray photoelectron spectroscopy (XPS) analysis on all the cation-intercalated membranes. The XPS survey of the K–V, Ca–V, and Al–V samples provided in Supplementary Fig. 9a shows common elements such as O, Si, Mg, and Al, along with intercalant atoms. The depth analysis of the intercalated cations indicates successful intercalation within the layers (Supplementary Fig. 9b-d). The O 1$s$ deconvoluted spectrum has two peaks for K–V, one at 531.0 eV and the other at 532.3 eV, which corresponds to the Si-O-Si(Al)/Si-O-Mg and hydroxyl groups, respectively (Fig. 1c)[34,35]. Both peaks shifted toward higher binding energy values with increasing cation valence, and the shift was maximal for Al-intercalated vermiculite, i.e., at 531.5 eV and 533.3 eV. The higher binding energy suggests that the electrostatic interaction between Al$^{3+}$ and the layers of vermiculite is considerable among all the intercalated vermiculite, which is in the order of Al$^{3+}$–V > Ca$^{2+}$–V > K$^{+}$–V. A similar shift is also reported in the case of cation-intercalated montmorillonite membranes[33]. A detailed analysis of the spectra corresponding to the elements of vermiculite, Si, Al, and Mg is provided in Supplementary Fig. 10.

We fabricated a voltage-gated device using various cation-exchanged vermiculite membranes. The membrane was initially cut into several pieces of size 5×4 mm$^2$. These membrane pieces were further utilized to make the device. For the gate voltage, a gold wire was fixed onto a silver strip made in the middle of the membrane, which ensures a uniform gate voltage across the membrane (Fig. 2). This sample was then encapsulated between two acrylic blocks with the help of an epoxy (Loctite Stycast 1266), which isolated the gold gate electrode from the electrolytes. To expose both ends of the membrane, the sample was polished with P1000 emery sandpaper, and the final dimensions were confirmed via optical microscopy. The polished sample was sandwiched between two circular acrylic pieces with a prefabricated hole to ensure that ion transport occurred only through the vermiculite membrane (Fig. 2). The fabricated device had a length, l of 3 mm, and width, w of 4 mm. We performed all our measurements with these devices.

We performed ionic transport measurements in a homemade PEEK (polyether ether ketone) cell. The device was placed between two reservoirs containing aqueous salt solutions. Two homemade Ag/AgCl electrodes, labeled as the source and drain, were used for the ionic measurement. The ions were transported by applying a voltage, $V_{ds}$ (drain to source voltage), and the resultant current, $I_{ds}$ (drain to source current), was monitored. We used another source meter to apply a gate voltage ($V_g$) to the nanofluidic device, with the source as the ground terminal. A schematic of the measurement setup is shown in the inset of Fig. 3a. For the transport studies, we choose the right salt-exchanged membrane and solutions, for example, K–V membranes with KCl solutions, Ca–V membranes with CaCl$_2$ and Al–V membranes with AlCl$_3$, to avoid any ion exchange during the experiments. We prepared all the salt solutions with Milli-Q deionized water, and the final pH of the solution was 5–7; however, AlCl$_3$ is an exception, with a pH of 2.7 for a 1 mM chloride concentration. The pH values of the Li$^+$, K$^+$, Ca$^{2+}$, and Al$^{3+}$-intercalated vermiculite flakes dispersed in a 1 mM chloride solution were 7.2, 6.8, 7.1, and 5.6, respectively. It is clear from various characterizations that small changes in pH do not affect d-spacing or stability since equilibrium is achieved due to cation exchange. The sample temperature was maintained at 298 K for all the measurements, unless otherwise specified.

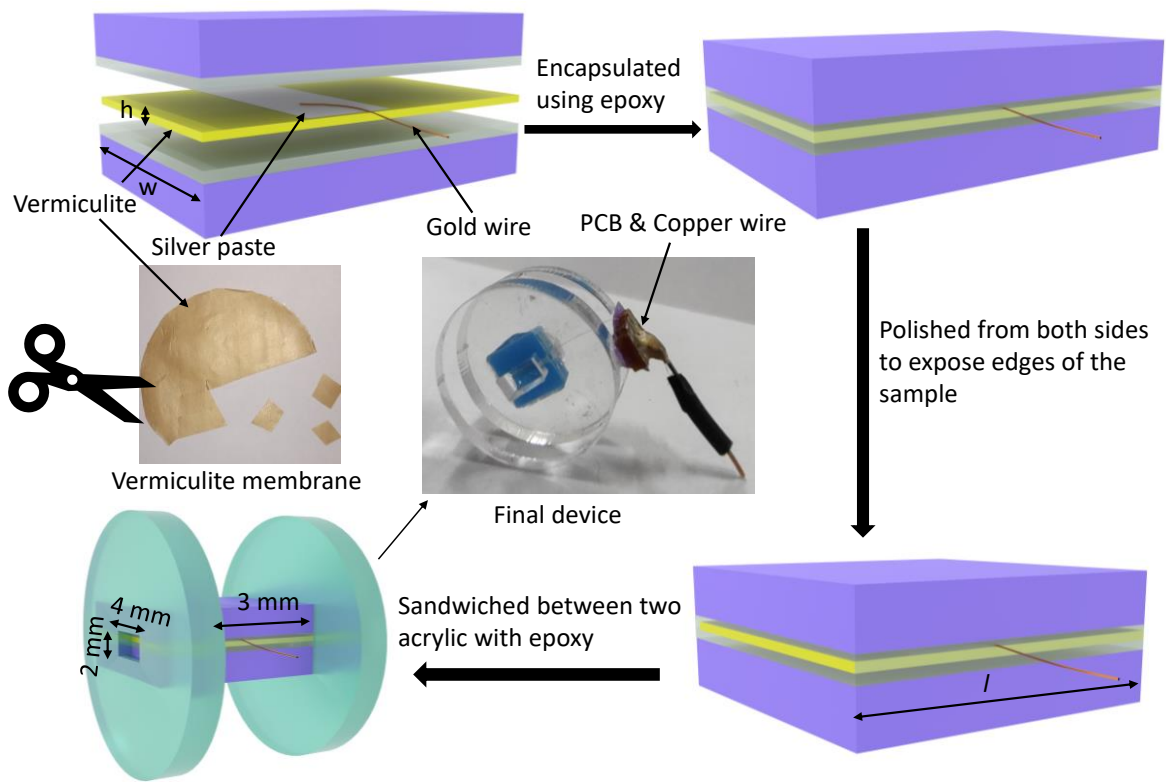

**Fig. 2 | Fabrication of a voltage-gated sample.** Schematics of various steps involved in the fabrication of voltage-gated vermiculite samples for studying gated ion transport characteristics. Here, h, w, and l are the thickness, width, and length of the membrane, respectively. We used a gold wire to apply the gate voltage. The images are not drawn to scale.

We first studied the effect of the gate voltage on the ion transport characteristics of a K–V sample. For this purpose, $I_{ds}$ - $V_{ds}$ characteristics were measured at different gate voltages, with $V_g$ ranging from -2 V to +1 V. The characteristics were mostly linear at low $V_{ds}$ (Fig. 3a) for a 1000 mM KCl solution. There is a visible change in the slope of the curve with various $V_g$ values. A finite zero-current voltage is observed for a larger negative $V_g$, plausibly indicating an asymmetric charge distribution. The larger positive current with negative $V_g$ is attributed to the increased $K^+$ density and mobility, as supported by MD simulations presented later. The ionic conductance was extracted from the slope of the $I_{ds}$–$V_{ds}$ plot. The KCl conductance across the membrane was reduced from 5.0 μS to 1.5 μS when $V_g$ increased to +1 V (Fig. 3b). Conversely, the conductance value increased to 22.9 μS after applying a gate voltage of –2 V. The conductance increased by 1400% when $V_g$ changed from +1 V to -2 V at a concentration of 1000 mM, which, to the best of our knowledge (comparison with other works is provided in Supplementary Tab. 1), is the highest ever reported (additional data can be found in Supplementary Fig. 11a&b). The conductance was highly reversible when $V_g$ was tuned back to +1 V from -2 V (Supplementary Fig. 12a&b), a clear indicator of the stability of our devices. Furthermore, at the end of each gate voltage cycle, we measured the water conductance to check the integrity of our devices. More than 90% of our devices were found to be stable in this gate voltage range.

To understand the role of the EDL in ion transport, we examined the magnitude of voltage modulation at several KCl concentrations in the range of 1 mM–1000 mM. For this purpose, we defined a gating ON/OFF ratio, which is the ratio of conductance at $V_g$ = -2 V to $V_g$ = +1 V (Fig. 3c). The gating ON/OFF ratio is smaller at a 1 mM concentration; however, it increases and saturates in the concentration range of 10 mM–1000 mM, which is very similar to the zero $V_g$ conductance versus concentration behavior (Supplementary Fig. 13). The increase in the gating ON/OFF ratio with increasing concentration is unusual and strikingly different from the findings of other reports[12,13]. It is therefore important to understand the reasons for the significant gate modulation effects at high KCl concentrations. For this, let us first consider the situation of zero gate voltage. The cations (K$^+$) from the KCl solution are drawn to the negatively charged vermiculite surface (Supplementary Note. 1) and are screened by the K$^+$ ions when the channel surface is not subjected to any gate voltage. The electrolyte solution concentration and ion valence control the screening length, also known as the Debye length, $\lambda_D$ , which can be computed via the following formula given as

$$\lambda_D = \sqrt{\frac{\varepsilon RT}{\sum_{i=1}^{N} F^2 Z_i^2 C_{i,0}}} \tag{1}$$

where $\varepsilon$ is the permittivity of the solution; $R$ and $T$ are the gas constant and temperature, respectively; $F$ is the Faraday constant; $N$ is the total number of ionic species; and $Z_i$ and $C_{i,0}$ are the valence and bulk concentrations of the $i^{th}$ ionic species, respectively. The above expression implies that the Debye length decreases with increasing ion concentration. For example, at a 1000 mM KCl concentration, $\lambda_D$ is 3 Å. The transport height for the K–V laminate is 2.6 Å, which means that for KCl concentrations ranging from 1 mM to 1000 mM, the EDL overlaps. Because of this, the variation in ionic conductance with concentration is not significant (Supplementary Fig. 13). For example, the conductance only slightly increased from 1.31 µS to 5 µS when the KCl concentration was varied from 1 mM to 1000 mM. This relatively small change in conductance, despite a three-orders-of-magnitude increase in concentrations, suggests that ion transport is dominated by surface effects rather than bulk diffusive transport, $G \propto C$ [Ref.[36]].

We tested whether the ion hydration shells are squeezed or the laminated membrane structure expanded upon an applied gate voltage ($V_g$). The in-situ XRD data (Supplementary Fig. 1b) confirm that the laminated structure remains intact after a high negative $V_g$ is applied. We tried to quantify the softness of the ion hydration shell, i.e., the extent of ion–water interactions and their modification under the influence of a gate voltage. We performed ionic conductance measurements at several temperatures and gate voltages to extract the activation energy ($E_a$). Fig. 3d displays the Arrhenius plot between ln($G$) and 1/$T$, at different $V_g$ values. $E_a$ is calculated using the following formula

$$G \propto \exp\left(\frac{-E_a}{RT}\right) \tag{2}$$

where $R$ is the universal gas constant, and the sample temperature $T$, varies from 25 to 60 °C. Our diffusion study revealed that K-V laminates are cation-selective due to the high density of negative surface charges (Supplementary Fig. 14 a&b & Supplementary Tab. 2). Therefore, we can readily ascribe the activation energy $E_a$ to that of K$^+$ ions. At 0 $V_g$, $E_a$ is 298 meV. When $V_g$ is +1 V, the $E_a$ value increases to 309 meV, which explains the reduced ionic conductance observed for positive $V_g$ (Fig. 3b). Interestingly, when $V_g$ = -2 V, $E_a$ decreased from 298 meV to 287 meV. The existence of a threshold voltage ($V_g \approx$ -1 V) and the lower $E_a$ (hence increased conductance) for $V_g$ < -1 V suggests a modification in ion–water (hydration) interactions around K$^+$ ions.

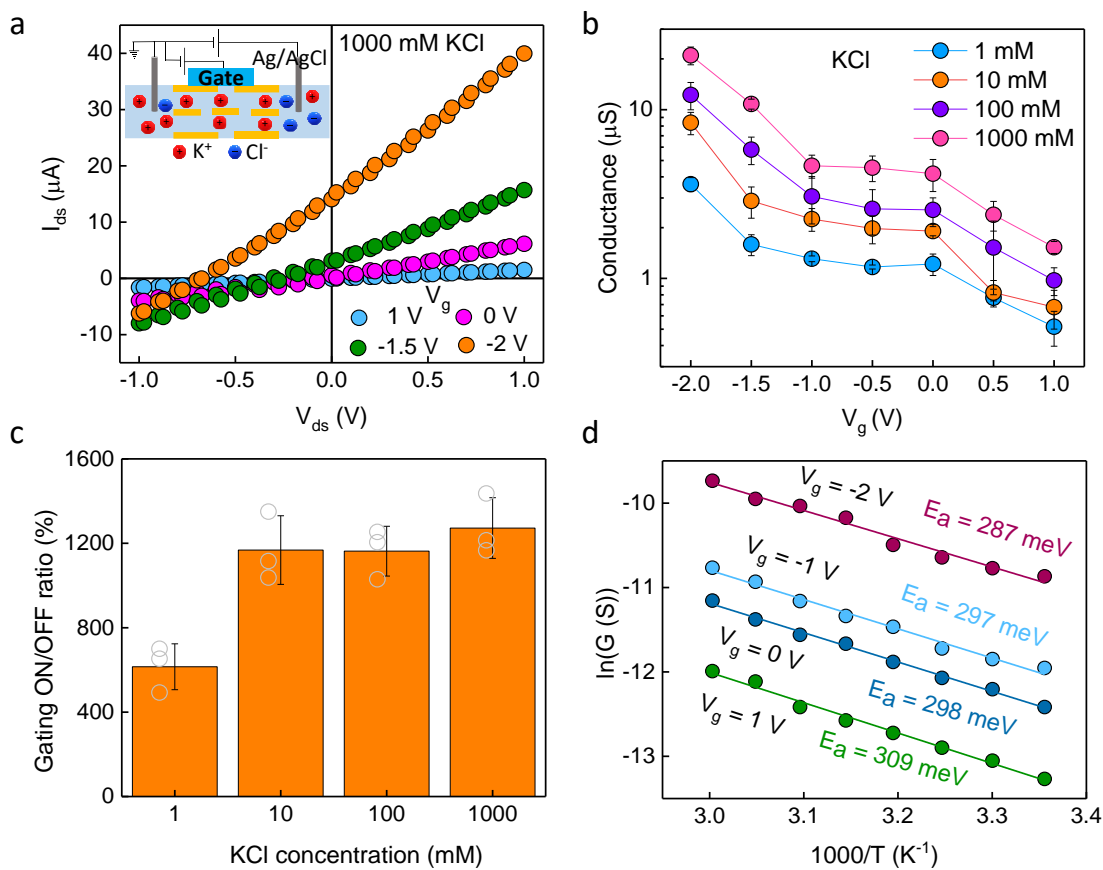

**Fig. 3 | Electric field modulated ion transport through the K-V membrane. a** Modulation of the ionic current, $I_{ds}$ with different $V_g$ values from -2 V to +1 V in steps of 0.5 V in 1000 mM KCl solution. Inset: schematic of the voltage-gated device used for the transport measurement. **b** Variation in conductance with gate potential for 1 mM, 10 mM, 100 mM, and 1000 mM KCl. The error bar indicates the standard deviations of 3 different samples. Data are presented as mean values +/- standard deviation. $N$ = 3 independent samples. **c** Gating ON/OFF ratio between conductance at -2 V to +1 V for the several concentrations of KCl from 1 to 1000 mM. Data are presented as mean values +/- standard deviation. $N$ = 3 independent samples. **d** Arrhenius plot for 1000 mM KCl showing the activation energy at several $V_g$ values, the solid lines represent a linear fit. Source data are provided as a Source Data file.

We examined the impact of the gate voltage on the transport of higher valency cations using the electrolytes $CaCl_2$ and $AlCl_3$, and the conductance vs. concentration characteristics at 0 $V_g$ are displayed in Supplementary Fig. 15a&b. Fig. 4a shows the *I–V* characteristics at different $V_g$ values for 500 mM $CaCl_2$ solutions. There is a visible change in the slope of the *I–V* curves with $V_g$. However, the current at zero voltage is lower than that of KCl solutions, especially for negative $V_g$. Fig. 4b shows the influence of the gate potential on the ionic conductance of $CaCl_2$ solutions at a concentration of 500 mM. For positive $V_g$, we observe a reduction in the $CaCl_2$ conductance upon an increase in $V_g$, similar to that of KCl solutions. However, for negative $V_g$, the conductance reaches a maximum at $V_g \approx$ -1 V, beyond which it decreases. We performed these measurements at several chloride concentrations ranging from 1 mM to 1000 mM (Fig. 4b). When we increased the solution concentration from 1 mM, the maximum conductance, which occurred at around $V_g \approx -1$ V for all the concentrations, also increased. We also wanted to confirm whether this observation holds for a trivalent cation such as $Al^{3+}$. For this purpose, we chose aqueous solutions of $AlCl_3$ with chloride concentrations ranging from 1 mM to 1000 mM. In this case, we also observed a maximum in the conductance, but at $V_g \approx 0$ V (Fig. 4c&d). The maximum conductance increased when we increased the solution concentration, a trend similar to that of $CaCl_2$ solutions.

The zero $V_g$ conductance of $CaCl_2$ and $AlCl_3$ is lower than that of KCl. Our estimated transport height for Ca-V and Al-V is 5.6 Å and 5.1 Å, respectively. It can accommodate 1-2 layers of water molecules as the bare sizes of $Ca^{2+}$ and $Al^{3+}$ ions are 2.02 Å and 1.06 Å, respectively[37]. Theoretical studies suggest that the first hydration layer is very strong in the case of $Ca^{2+}$ and $Al^{3+}$, which is difficult to remove as penalties of -1504 kJ/mol and -4665 kJ/mol energy are needed, respectively[37]. Therefore, major modification in the ion–water interaction is unlikely even with the largest $V_g$ that we have applied. Like in KCl solutions, the $CaCl_2$ and $AlCl_3$ conductance saturates at high concentrations (Supplementary Fig. 15a&b). Contrary to the KCl results, we observed a decrease in conductance with negative $V_g$, which suggests the emergence of a different interaction. Moreover, the maximum conductance decreases with decreasing ion concentration. Our diffusion data (Supplementary Fig. 14 a&b) and zeta potential measurements (Supplementary Fig. 5) indicate that the cation selectivity of vermiculite is in the order of KCl > $CaCl_2$ > $AlCl_3$.

We observed that the gate voltage dependent ionic conductance behavior of vermiculite laminates that are intercalated with KCl, $CaCl_2$, and $AlCl_3$ solutions is strikingly different from other reported membranes that are unintercalated, such as on MXenes[13]. The possibility of water splitting at extreme values of $V_g$ is ruled out simply because the conductance of $CaCl_2$ and $AlCl_3$ solutions decreases with $V_g$ beyond -1 V. In the case of $K^+$ ion transport through K–V laminates, we see a nonmonotonic increase in conductance with negative $V_g$. Recently, an atomic transistor consisting of rGO showed an increase in $K^+$ transport when a gate voltage was applied[14]. In this case, the density of ions increases, and the barrier energy for ions decreases with $V_g$, which results in increased conductance. The barrier energy is exponentially related to the ion mobility ($\mu$) as

$$\mu = \mu_0 \, exp\left(\frac{-E_a}{RT}\right) \qquad (3)$$

where $\mu_0$ is the bulk ion mobility of the ion (i.e. for $K^+$, $\mu_0$ = 7.62×10$^{-8}$ m$^2$V$^{-1}$s$^{-1}$). A decrease in the barrier energy clearly leads to an increase in ion mobility and hence in the conductance.

We also checked whether the gating ON/OFF ratio depends on the cation's energy barrier. For this purpose, we chose a Na–V membrane with NaCl solutions at a concentration of 1000 mM. We found that the gating ON/OFF ratio decreased to ~1100% (Supplementary Fig. 16a&b), which could be due to its ($Na^+$) slightly larger hydration barrier than $K^+$. Additionally, with increasing membrane thickness (K–V, h = 8 μm), the overall ionic conductance increases, but the gating ON/OFF ratio decreases to ~1100% (Supplementary Fig. 17a&b). This result suggests that the gating effect might be more pronounced in thinner than in thicker membranes.

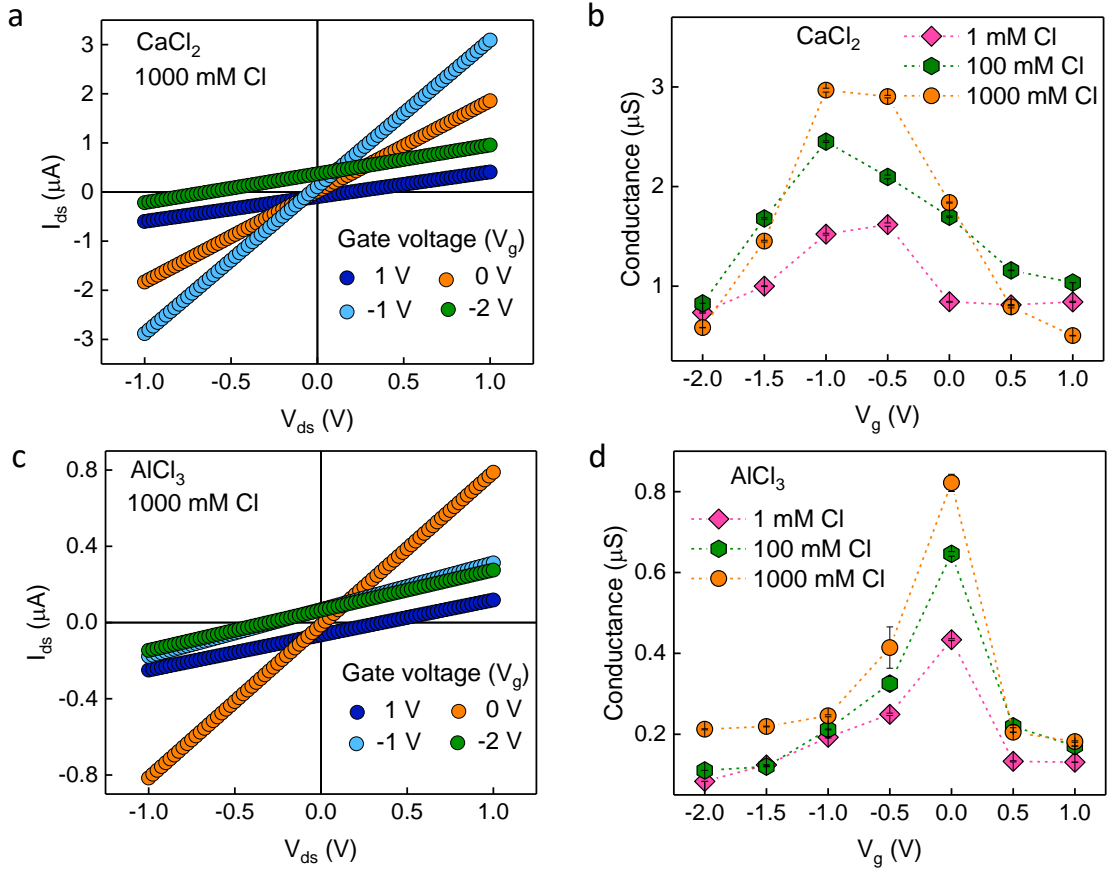

**Fig. 4 | Voltage-gated ion transport through Ca–V and Al–V devices, and the effect of ion valence. a** Variation in the ionic current, $I_{ds}$ with the applied potential, $V_{ds}$, for 500 mM $CaCl_2$ solutions at various $V_g$. **b** Conductance vs $V_g$ plot for 0.5 mM, 50 mM, and 500 mM $CaCl_2$ solutions. The error bar indicates the standard deviation for three independent measurements. Data are presented as mean values +/- standard deviation. $N = 3$ independent measurements. **c** Variation in the ionic current, $I_{ds}$ with the applied potential, $V_{ds}$, for 333 mM $AlCl_3$ solutions at various $V_g$. **d** Conductance vs $V_g$ plot for 0.33 mM, 33 mM, and 333 mM $AlCl_3$ solutions. The error bar indicates the standard deviation for three independent measurements. Data are presented as mean values +/- standard deviation. $N = 3$ independent measurements. Source data are provided as a Source Data file.

To understand the ion transport mechanism under gate voltages for K-V and Ca-V systems, we conducted all-atom MD simulations. The MD simulation setup is illustrated in Fig. 5a. We considered transport heights of 2.6 Å and 5.6 Å for the 1000 mM KCl and 500 mM $CaCl_2$ systems, respectively, matching the experimental values. To emulate the gating effect, a net charge $\Delta Q_{\text{net}}$ was introduced to the membrane atoms[38] using the distribution equation as

$$\Delta q_i = \Delta Q_{\text{net}} \frac{|q_i|}{\sum_i |q_i|} \quad (4)$$

where $q_i$ represents the partial charge of atom $i$ in the membrane. The charge modification applied was within 1.5% of the original partial charge. Ideally, the surface charge density $\sigma$ has a linear relation with $V_g$ which is given below

$$\sigma = \frac{\Delta Q_{\text{net}} + Q_0}{A} = CV_g + \sigma_0 \tag{5}$$

where $A$ is the surface area, $C$ is the capacitance, and $Q_0$ and $\sigma_0$ are the net charge of the membrane and the surface charge density at $V_g = 0$, respectively. An external electric field corresponding to 1 V was applied to induce ion transport. Additional methodological details of the MD simulations are provided in the Methods section.

The surface of the vermiculite membrane is terminated with oxygen atoms, naturally generating negative electrostatic dipoles directed toward the interlayer space, thereby providing binding sites for cations. This locally negative electrostatic environment within the highly confined vermiculite interlayer attracts cations, even when the membrane is electrically neutral. Fig. 5b shows snapshots of ions in K-V and Ca-V systems, demonstrating that a negative gate voltage generally increases the population of intercalated cations, whereas a positive gate voltage decreases it. Note that cation transport through this highly confined space involves a sequential hopping process between binding sites created by the oxygen-terminated surface. Such a sequential, barrier-leaping ion transport mechanism has been historically studied in the context of biological nanochannels[39] and has also been documented in synthetic membranes[40]. Our MD simulations for cation currents in the K-V and Ca-V systems show qualitatively consistent trends with the experimental observations (Fig. 5c-d). In the K-V system, the cation current increases with decreasing σ (decreasing $V_g$), whereas the Ca-V system exhibits a peak current at a specific σ value. We calculated average ion density within the interlayer space as a function of surface charge density (see Fig. 5e–f). In the K-V system (2.6 Å), only K⁺ ions intercalate, whereas in Ca-V systems (5.6 Å), both Ca²⁺ and Cl⁻ ions occupy the interlayer space. Fig. 5g–h shows the average mobility of ions between the membranes. In the K-V system, K⁺ mobility increases with decreasing surface charge. In the Ca-V system, the Ca²⁺ mobility peaks at σ ≈ -0.4 C/m², whereas the Cl⁻ mobility decreases with decreasing σ.

We closely examined ion trajectories at different gate voltages in our MD simulation, leading to the following insights into ion transport mechanisms (illustrated in Fig. 5b): For the K-V system at negative $V_g$, the density of K⁺ increases and K⁺ ions undergo frequent, short-distance hopping, resulting in high current (Supplementary Movie 1). At a positive $V_g$, K⁺ ions become depleted and exhibit infrequent hopping over relatively long distances, leading to a low current (Supplementary Movie 2). For the Ca-V system, Ca²⁺ ions become less mobile at strongly negative $V_g$ because of the strong Ca-V interaction (Supplementary Movie 3). At positive $V_g$, Ca²⁺, and Cl⁻ ions move in opposite directions, interfacing with each other's transport and resulting in a congested hopping process that reduces Ca²⁺ mobility (Supplementary Movie 4). As a result of reduced mobility at both the negative and positive $V_g$ limits, the Ca²⁺ current reaches its maximum at the intermediate $V_g$ value (Supplementary Movie 5). Theoretically, the mobility of K⁺ ions can also be reduced at even higher negative surface charges. To further understand the effect of gating on ionic mobility within these Å scale channels, an energetic analysis of each ion may be necessary, although this is beyond the scope of this work.

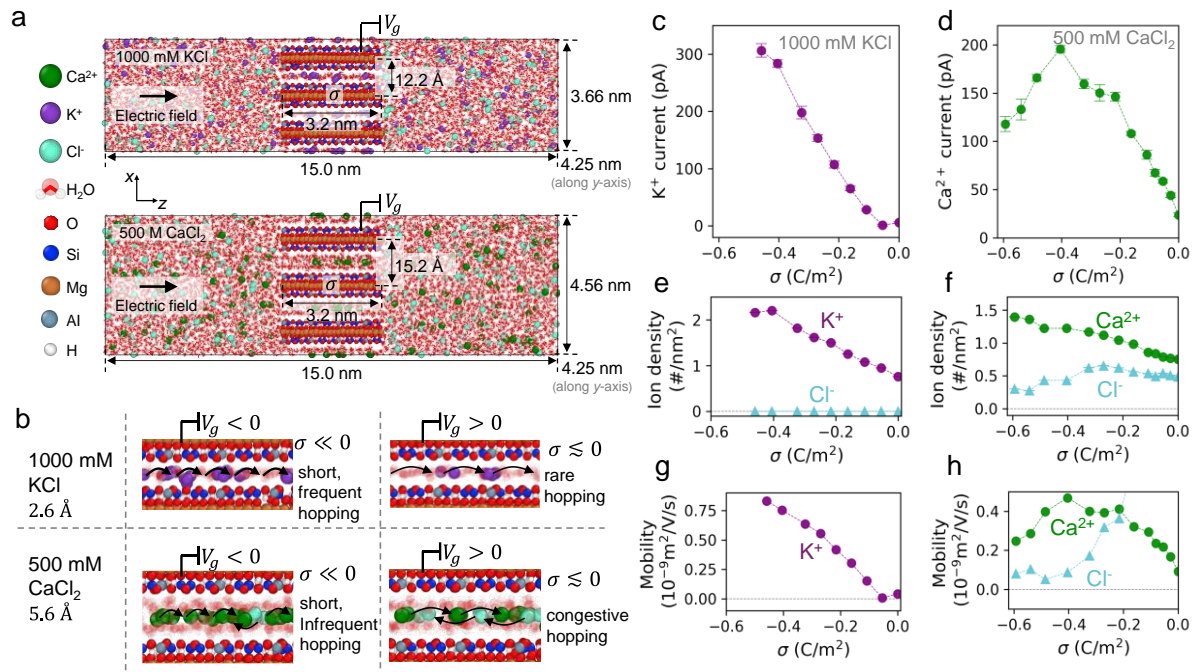

**Fig. 5 | Molecular dynamics (MD) simulation of ion transport through Å-scale 2D vermiculite membranes. a** Snapshot of the simulation systems for 1000 mM KCl (top) and 500 mM CaCl₂ (bottom) solutions. **b** Effect of the gating voltage on the ion distribution and transport through the vermiculite interlayer spacing. **c-d** Cation current as a function of surface charge density for **c** 1000 mM KCl and **d** 500 mM CaCl₂. The error bars for the current were estimated as the standard deviation of the slope obtained from a linear regression of the flux-time data, sampled every 50 ps over a total simulation time of 200 ns. The centre of the error bar is the mean. **e-f** Density of intercalated ion as a function of surface charge density for **e** 1000 mM KCl and **f** 500 mM CaCl₂. The error bars for the ion density were estimated as the standard deviation of the ion density, sampled every 50 ps over a total simulation time of 200 ns. The centre of the error bar is the mean. **g-h** Mobility of intercalated ions as a function of surface charge density for **g** 1000 mM KCl and **h** 500 mM CaCl₂. Source data are provided as a Source Data file.

We also investigated a commonly used GO system to infer the gating effect at 1000 mM, where the Debye length does not overlap with the transport height (~7 Å). We found no evidence of a gating effect (Fig. S18a–d).

The significant electric field effect observed in vermiculite membranes is a result of (i) the considerable length of the gate electrodes (~2 mm), which enhances cation accumulation when negative gate voltages are applied, and (ii) the ~1 nm thick insulating vermiculite layers capacitively induce charges, and the presence of 1 to 2 layers of water ensures weaker electrostatic screening, resulting in a field effect on the fluidic channels. Theoretical simulations indicate that thinner dielectric layers induce higher surface charge and zeta potential[41]. Recent reports of the electric field effect in conducting membranes such as GO and MXenes without the use of a gate oxide for gating are puzzling; however, it may be argued that there are thin insulating layers in the form of functional groups, which might help induce surface charges and hence tunability of conductance or permeation. Our device configuration is inspired by solid-state field-effect transistors; thus, the interpretation of the results is straightforward. Mica has a very high breakdown strength of 12 MV/cm [ref.43], which is expected to be similar for vermiculite. Therefore, the probability of the formation of conducting filaments in our membranes is very unlikely since our maximum applied voltage is only 2 V across a membrane thickness of 3.5 µm.

The gated ion transport device demonstrated in our study is effective in modulating the conductance of monovalent ions and, as such, holds promise for several advanced applications. First, the precise control over ion flow in angstrom-scale channels makes these devices excellent candidates for lab-on-chip systems, where accurate dosing and real-time regulation of ionic species are critical. For instance, in controlled drug delivery, the ability to selectively modulate the transport of monovalent ions could be leveraged to trigger downstream chemical or biological processes with high temporal resolution. Additionally, the distinct response observed for monovalent *vs.* multivalent ions – evidenced by the contrasting transconductance behavior under negative gate voltages – suggests that these devices could be engineered to separate ions based on their valence. This selective ion separation could have significant implications in water purification, where the removal of specific ion types is desired, as well as in the design of fuel cells and batteries. In these energy applications, energy-efficient ion gating is essential to optimize performance and reduce losses. The inherent modularity of the vermiculite membrane system allows for further functionalization. For example, by incorporating inorganic pillars (such as alumina) to expand the van der Waals gap, the device architecture could be tuned to accommodate larger species[42]. This would enable the transport of biomolecules (*e.g.*, DNA, RNA, proteins), finding applications in biosensing and gate-controlled sequencing and detection platforms.

We demonstrated how gate voltage alters ion-water and ion-surface interactions at room temperature in highly confined vermiculite laminates. Negative gate voltages increased the cation concentration inside the channels, resulting in unusual ion gating effects. We observed a sizeable ionic conductance modulation of ∼1400% in 1000 mM KCl solutions. In the same voltage range, much-reduced conductance modulation was observed in the $CaCl_2$ and $AlCl_3$ solutions, resulting from ion-surface interactions inferred from the MD simulations. Future research must focus on high salt concentrations and extreme confinements to understand and explore the whole dynamics of ion interaction effects. This study reveals how to treat industrial waste of high salt concentrations efficiently. Overall, two-dimensional materials offer endless possibilities for fabricating tunable channels at the Å-scale.

**Methods**

**Chemicals:** Natural vermiculite crystals (2–3 mm in size), potassium chloride (KCl, ≥ 98.5%), sodium chloride (NaCl, ≥ 99.0%), lithium chloride (LiCl, ≥99.0%), calcium chloride ($CaCl_2$, ≥98.0%), aluminum chloride ($AlCl_3$, ≥97.0%), PVDF (0.22 µm pore size), and silver paste were purchased from Sigma Aldrich. All the chemicals purchased were used as received. Gold wire (30 µm in diameter) was purchased from Tanaka K.K. (Japan).

**Fabrication of vermiculite laminates:** The natural vermiculite crystals obtained were thermally expanded, and the cations were exchanged via a two-step method[25]. Briefly, in the first step, 100 mg of vermiculite crystals were soaked in a 200 mL saturated NaCl solution and refluxed for 24 h at 100 °C, followed by washing with deionized water 8 to 10 times to remove excess salts. During this process, interlayer $Mg^{2+}$ cations are exchanged with $Na^+$ ions. In the next step, the sodium-exchanged vermiculites were soaked in 200 mL of 2000 mM lithium chloride (LiCl) solution, refluxed for an additional 24 h, and again washed with DI water until excess chloride ions were removed. The resulting Li-exchanged vermiculite (Li-V) crystals were dried and dispersed in water at a 1 mg/mL concentration and sonicated for 30 minutes. The monolayers were obtained by centrifuging at 1509×g for 15 minutes or allowing the large flakes to settle in the solution overnight under gravity. We used the supernatant containing the monolayer to prepare the vermiculite membrane via vacuum filtration assembly. We used PVDF of 0.22 µm pore size as the support for vermiculite membranes, which easily peeled off after drying the samples under an IR lamp for 10 min. These free-standing membranes were dipped into 1000 mM chloride concentration of KCl, $CaCl_2$, and $AlCl_3$ solutions for 24 hours, which resulted in the intercalation of the cations $K^+$, $Ca^{2+}$, and $Al^{3+}$, respectively, into the vermiculite interlayers. Our previous study[25] reported that the above cations make vermiculite membranes highly water-stable. We subsequently washed the cation exchanged membranes with water to remove excess salt from the surface and dried them under an IR lamp. These materials were further characterized and used for ion transport studies.

**Characterization:** A Rigaku Multipurpose X-ray diffractometer (XRD) with Cu Kα radiation (λ = 1.5406 Å) was used to determine the interlayer spacing of the cation-exchanged vermiculite membranes. The thickness of the exfoliated vermiculite flake was determined using atomic force microscopy (Bruker Nano wizard Sense AFM), the membrane thickness was determined from cross-sectional SEM images (JEOL JSM-7900F), and a zeta potential measurement was carried out via DLS and Zeta Seizer instrument (Nano ZS Malvern Instrument). Energy dispersive analysis (EDS, JEOL JSM-7900F) was performed to determine the elements present in the vermiculite membranes. Fourier transform infrared spectroscopy (FTIR) (Perkin Elmer), and Raman spectroscopy (WITec) were used to confirm the functional groups present in vermiculite. For Raman spectroscopy, 532 nm laser light with 5 mW of power was used for excitation at room temperature. X-ray photoelectron spectroscopy (XPS) was also carried out to determine the elemental details. XPS spectra were acquired with an ESCALAB 250 XI (Thermo Fisher Scientific, source: Mg Kα 300 W, pass energy: 40 eV) system, where the analysis chamber was pumped down to ultrahigh vacuum (UHV ~5 x $10^{-10}$ mbar). The mechanical testing was carried out using a universal testing machine (UTM, model number- Kappa SS_CF100).

**Molecular dynamics simulation:** All-atom molecular dynamics (MD) simulations were performed to investigate ion transport through Å-scale two-dimensional (2D) vermiculite membranes. The simulations were conducted using periodic rectangular simulation boxes with dimensions of 3.66 nm × 4.25 nm × 15.0 nm (x, y, z) for the 1000 mM KCl system and 4.56 nm × 4.25 nm × 15.0 nm for the 500 mM $CaCl_2$ system. Three layers of vermiculite membranes were stacked along the x-direction and positioned at the center of the z-coordinate system, as depicted in Fig. 5a. The vermiculite membranes and ions were modelled using the Clay Force Field[44] and water molecules were described by the standard simple point charge (SPC) model[45]. To emulate the gating effect of the membranes, a net charge ($Q_{\text{net}}$) was added to the membrane atoms using the distribution equation described in the main text. The systems were solvated with water molecules and ions with additional cations added to neutralize the system. The positions of the vermiculite membrane atoms were restrained using a harmonic spring. Prior to production runs, the system underwent an initial energy minimization, followed by equilibration under the NPT ensemble with a variable z-box size at 300 K and 1 bar for 2 ns. The barostat and thermostat relaxation times were set to 2 ps and 0.1 ps, respectively. Following equilibration, an external electric field of 0.067 V/nm was applied along the z-direction, corresponding to a 1 V transmembrane bias, and the simulation was conducted under the NVT ensemble. Ion transport simulations were performed for 200 ns, and the last 100 ns of trajectory data were used for analysis. All MD simulations were performed using a GPU-accelerated MD code with GROMACS[46,47] version 2024.2, and atomic visualizations were generated using OVITO[48].

**Data Availability**

All data that are required to understand the conclusions in the paper are presented in the main manuscript and the supplementary information.

**Acknowledgments**

This work was mainly funded by DST-INAE with grant no. 2023/IN-TW/09. We also acknowledge the financial support from MHRD STARS with grant no. MoE-STARS/STARS-1/405 and by Science and Engineering Research Board (SERB), Government of India, through grant no. CRG/2023/004818. K.G. acknowledges the support of Kanchan and Harilal Doshi chair fund. C.N. acknowledges the support from IASc-INSA-NASI in the form of Focus Area Science Technology Summer Fellowship. The authors acknowledge the contribution from IITGN central instrumentation facility, especially the FIST-DST system with grant no. SR/FST/PS-1/2020/141. L.H.Y. acknowledges the financial support from the National Science and Technology Council (NSTC), Taiwan under Grant No. NSTC 112-2923-E-011-003-MY. Use of computational resources was supported by NSF under the ACCESS program, award no. PHY250014, to Y.N.

**Author Contributions Statement**

K.G. conceived the idea and supervised the project. D.B. designed the project, executed the sample preparation, characterization, voltage gating measurements, and result analysis. Y.N carried out the MD simulation and result analysis. S.N.P helped in the result analysis. C.N. helped in the sample preparation and initial optimization of the voltage-gating measurements. S.S.N. carried out the Raman and XPS measurements. R.A. helped in optimizing the membrane stability and zeta potential measurements. K.S. helped in the XRD measurement. A.C. and L.H.Y. helped in the data analysis. S.K.N. helped in the theoretical analysis. D.B., Y.N, and K.G. wrote the manuscript. All the authors discussed the results and commented on the manuscript.

**Competing Interests Statement**

The authors declare no competing interests.

# Supplementary information

**Contents:**

**Supplementary Figures:**

Supplementary Fig. 18. Characterization and gated ion transport studies through the GO sample.

**Supplementary notes:**

Supplementary note 1. Surface charge density calculation.

Supplementary note 2. Diffusion studies.

Supplementary note 3. Fabrication of graphene oxide membrane.

**Supplementary tables:**

Supplementary table 1. Summary of gating ON/OFF ratio for different gated nanofluidic systems.

Supplementary table 2. Activity coefficients and ion selectivity of K–V membranes in KCl solutions.

**References 1 - 8**

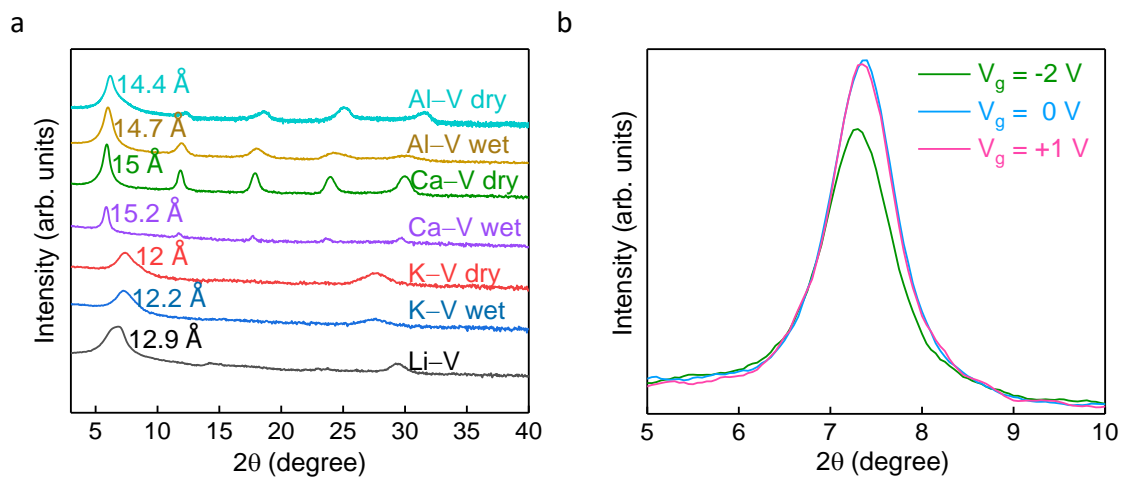

**Supplementary Figure 1. X-ray diffraction (XRD) analysis of the cation-exchanged vermiculite membranes. a** X-ray diffraction pattern of the Li–V, K–V, Ca–V, and Al–V membranes in both dry and wet states. The interlayer spacing is ~12 Å for K-V and Li-V membranes, and for Ca-V and Al–V membranes, it is ~15 Å. **b** In-situ X-ray diffraction pattern of the K–V device with various applied $V_g$ at a solution concentration of 1000 mM KCl. We varied $V_g$ from -2 V to 1 V, and a solution of 1000 mM KCl concentration was used, which mimics our transport measurement conditions. It is observed from the XRD data that when $V_g$ = 0 V, the intense peak of K–V is at 7.38° which changes to 7.29° at a $V_g$ of -2 V. The estimated d-spacing of K–V membrane changes very little from 12 Å to 12.1 Å, a net change of 0.1 Å. It confirms that the interlayer spacing is not affected by gate voltage.

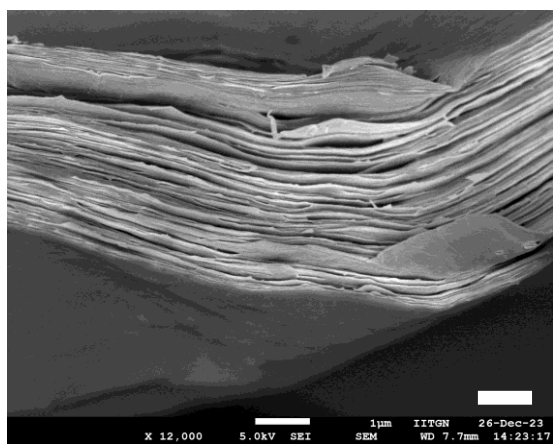

**Supplementary Figure 2. Scanning electron microscopy (SEM) characterization of vermiculite membrane.** Cross-sectional image of the K–V membrane shows laminated structure with thickness ~3.5 μm. The scale bar is 1 μm.

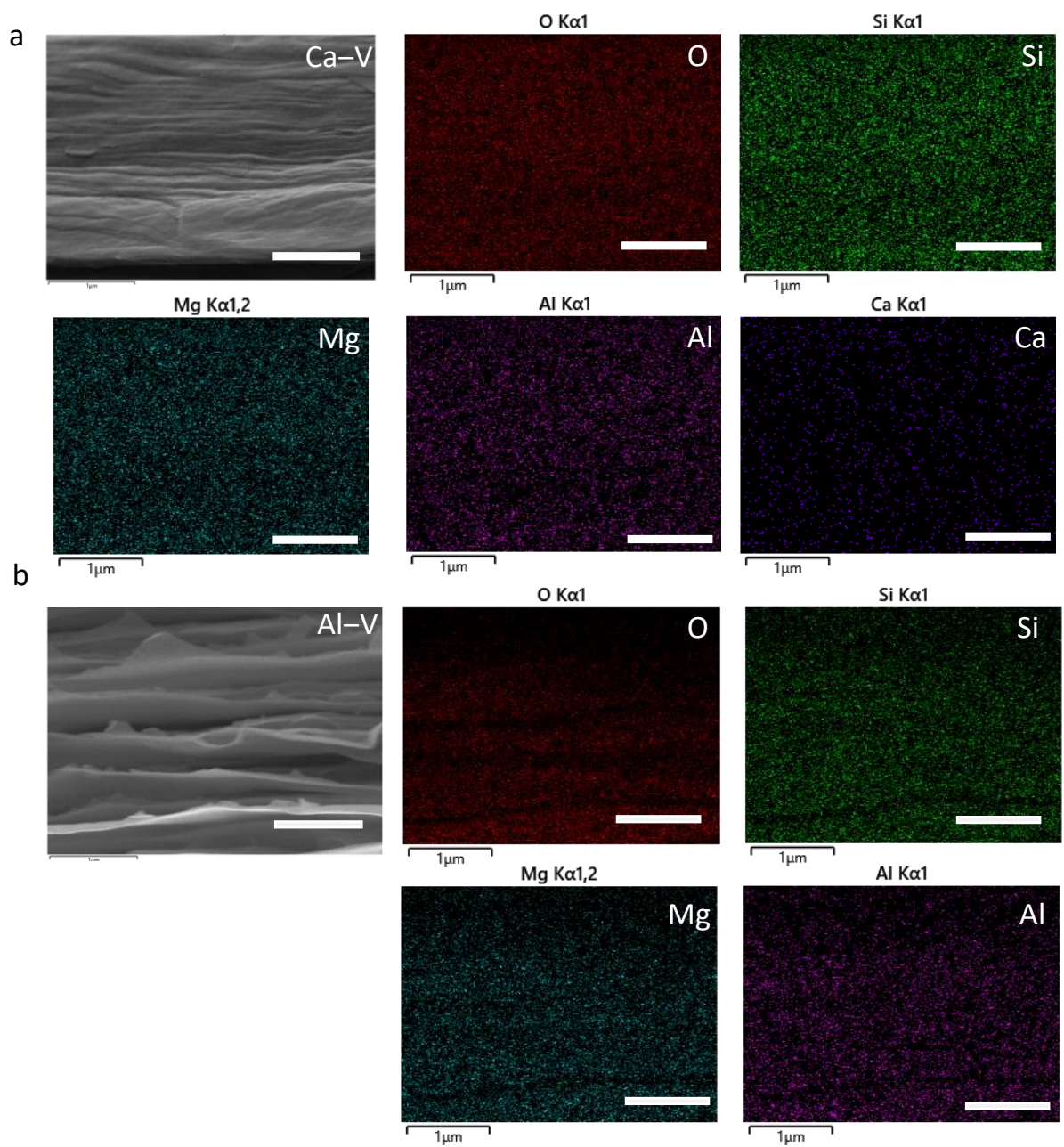

**Supplementary Figure 3. Cross-sectional energy dispersive (EDS) elemental mapping of Ca–V and Al–V membranes.** The scale bar represents 1 μm. From the EDS mapping, we found that the main elements present are O, Si, Mg, and Al. **a** For Ca–V membranes, intercalant Ca is also detected. **b** In Al–V membranes, the signal arising from the intercalant Al is merged with the signal from the intrinsic Al, so no separate detection is possible.

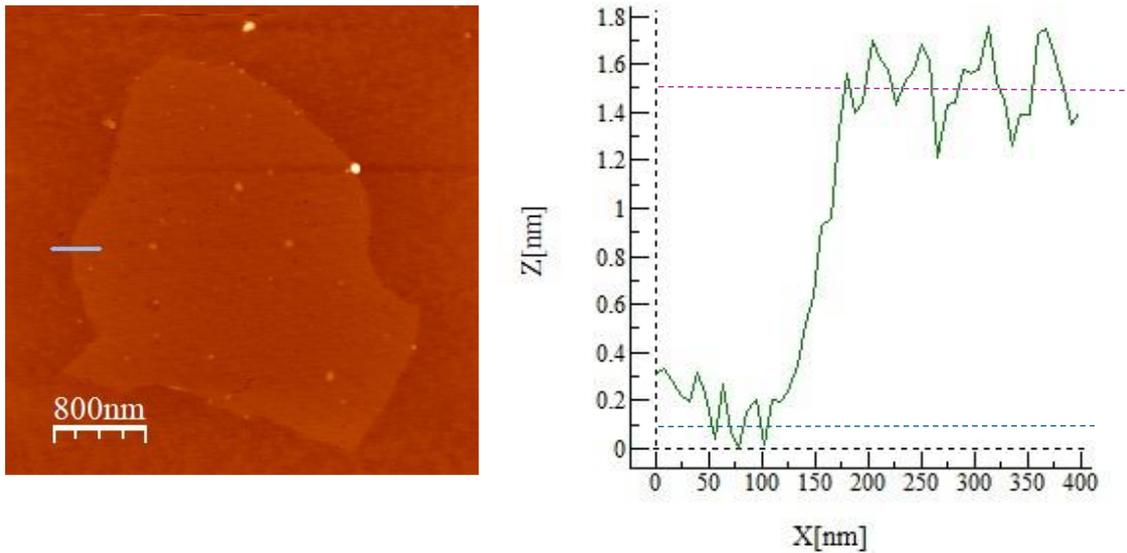

**Supplementary Figure 4. Atomic force microscopy (AFM) image of the vermiculite.** AFM image of monolayer vermiculite laminate with a flake size of ~2 µm. The flake sizes are typically in the range 1-3 µm. The height profile (the right image) provides a thickness of 14 Å for single-layer vermiculite. Mica-like structure adsorbs 1 or 2 layers of water molecules on its surface and also hosts intercalated ions on its surface. This results in a slightly larger thickness of 14 Å instead of 9.6 Å.

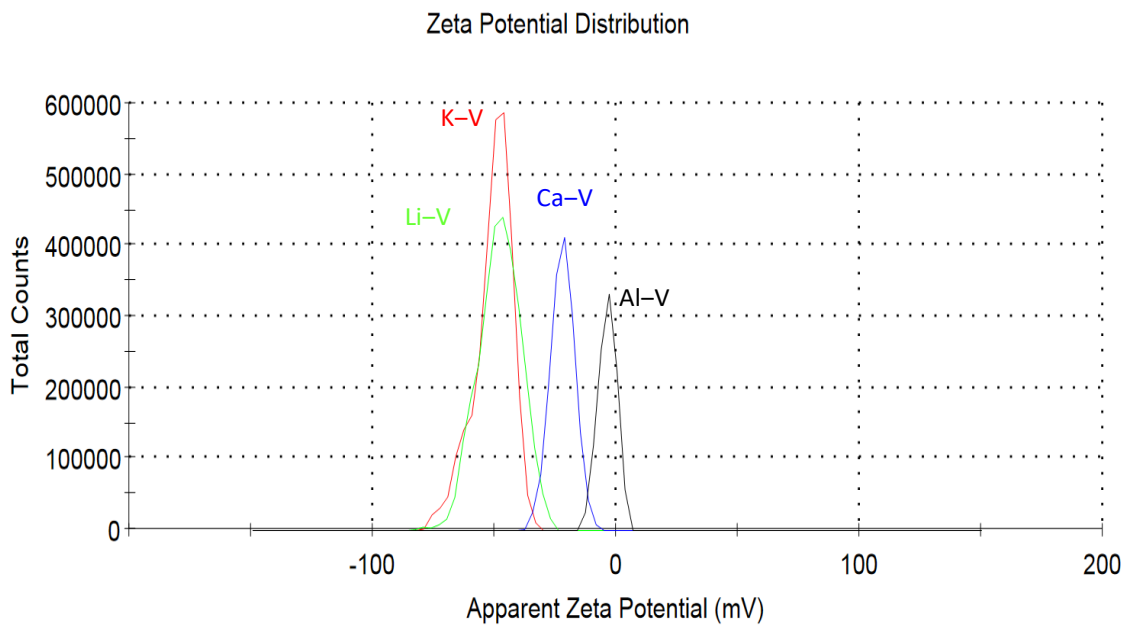

**Supplementary Figure 5. Zeta potential characterization of cation-exchanged vermiculite laminates.** Zeta potential measurement of the dispersed vermiculite flake solution with $Li^+$, $K^+$, $Ca^{2+}$, and $Al^{3+}$ intercalant is displayed. The concentration of the dispersed solution was 1 mg/mL in deionized water (DI). We measured the zeta potential in the membrane as well as in dispersed forms, and in both cases, the zeta potential is found to be similar. With an increase in valence of the exchanged cation in vermiculite, the zeta potential is found to decrease.

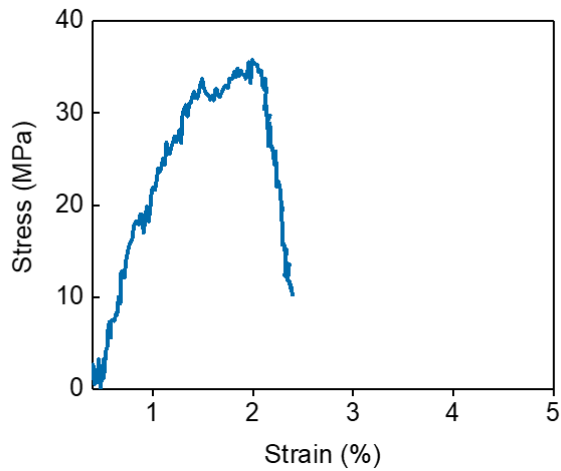

**Supplementary Figure 6. Mechanical strength analysis of the vermiculite membrane.** The tensile strength of vermiculite is approximately 35 MPa, with a fracture strain of about 2%. This indicates that vermiculite membranes are flexible and suitable for use in aqueous conditions.

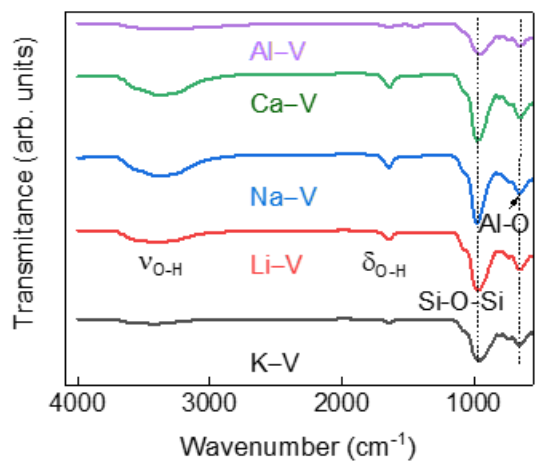

**Supplementary Figure 7. FTIR analysis of the different intercalated membranes.** The broad absorption peak at 3360 cm$^{-1}$ and 1643 cm$^{-1}$ indicates stretching vibration and bending vibrations of the -OH groups. The absorbance peak at 970 cm$^{-1}$ is ascribed to the asymmetric stretching vibration of the Si-O-Si. Two peaks at 735 cm$^{-1}$ and 652 cm$^{-1}$ can be attributed to the Al-O-Al and M-O-Si (Mg, Al, Fe) bonds, respectively.

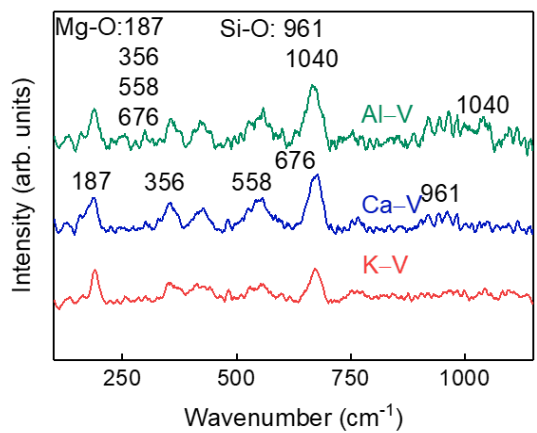

**Supplementary Figure 8. Raman spectra of the K–V, Ca–V and Al–V membranes**. The spectra show the characteristic peak of Mg-O/Al-O at positions 187, 356, 558, and 676 cm$^{-1}$ and Si-O at 961 cm$^{-1}$ and 1040 cm$^{-1}$, thus confirming the compositional aspect of magnesium aluminosilicate.

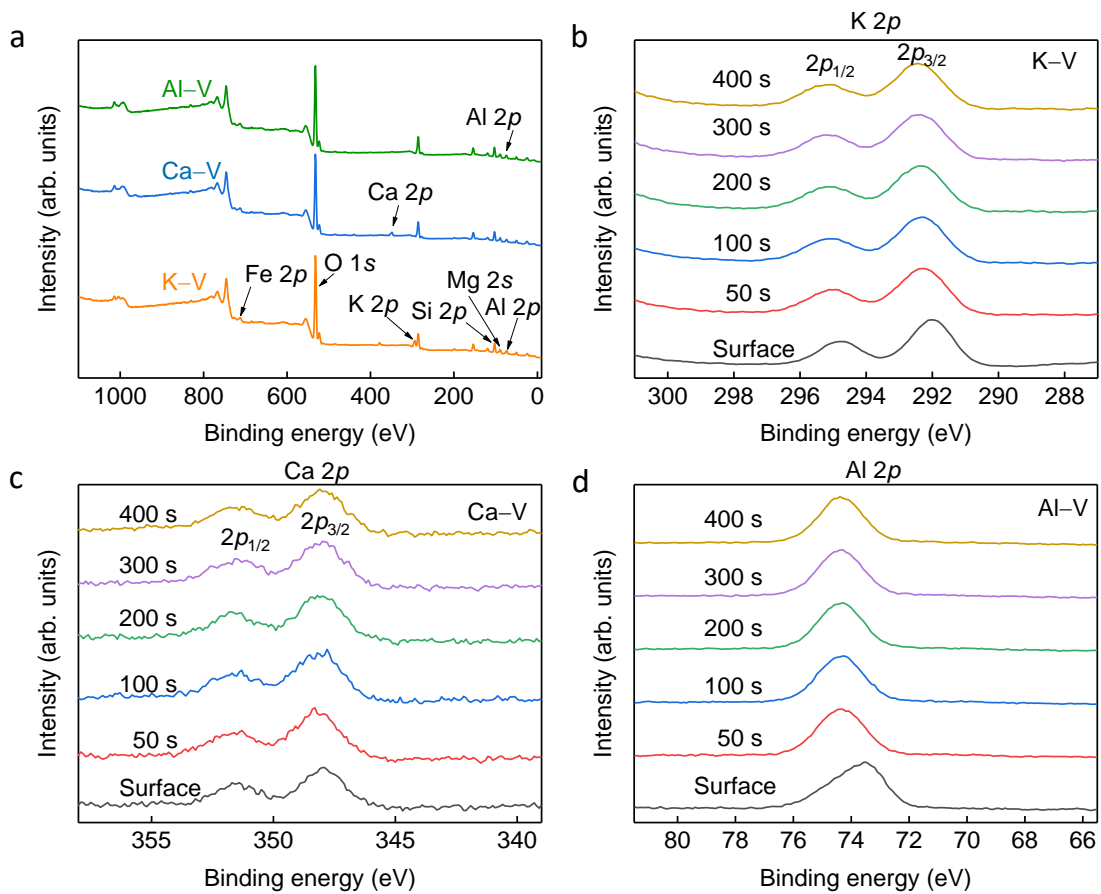

**Supplementary Figure 9. XPS survey scan and depth analysis of the cation-intercalated vermiculite membranes. a** XPS survey spectrum of K–V, Ca–V, and Al–V membranes. It displays common elements present in the intercalated membrane, such as O, Si, Mg, and Al. Additionally, K, Ca, and Al (merge with the intrinsic Al) signals are detected from K–V, Ca–V, and Al–V membranes, respectively, which confirms the successful intercalation of the cations inside vermiculite. **b** Depth profile scan of K–V membrane indicates the presence of K inside deeper layers of the membrane with uniform distribution. **c** Depth profile scan of Ca–V membrane confirming the presence of Ca inside the layers and its uniform distribution. **d** Depth profile scan of Al–V membrane indicates the presence of Al across the membrane, and Al is uniformly distributed. Note that we could not distinguish the intercalated Al ion and the Al present in the tetrahedral sites.

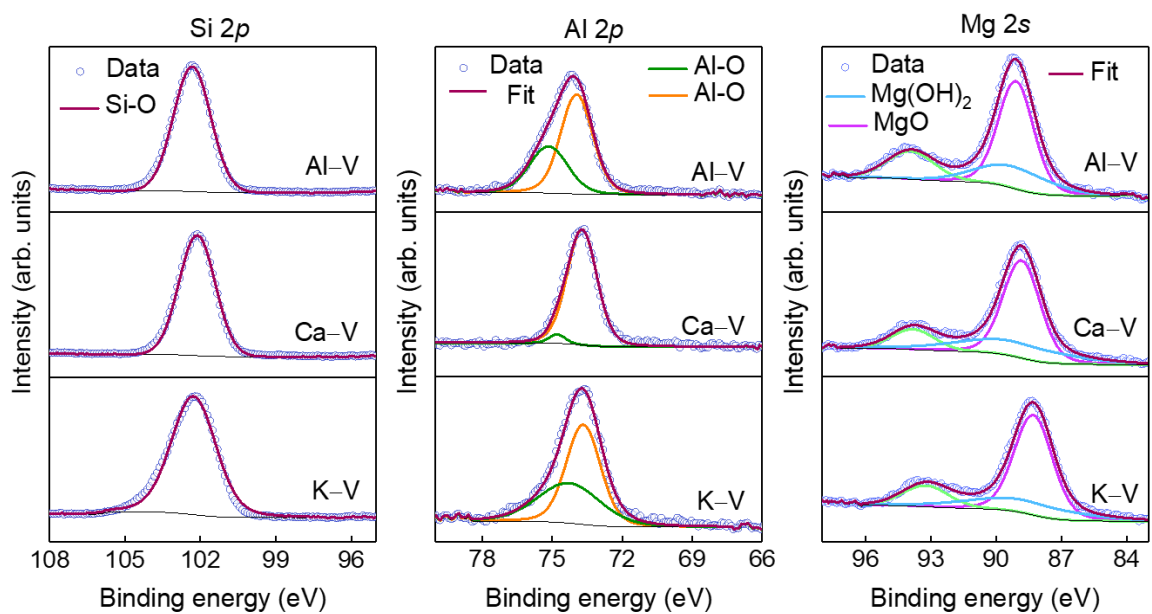

**Supplementary Figure 10. X-ray photoelectron spectroscopy (XPS) studies of intrinsic elements of vermiculite, Si, Al, and Mg (O 1s is given in the main text).** The Si 2p spectrum shows a peak at 102.3 eV for K–V and Ca–V membranes corresponding to Si-O. For Al–V membranes, the Si 2p peak shows a slight shift towards higher binding energy side when compared to K–V (or Ca–V) membranes, which could be a result of increased ionicity due to intercalant $Al^{3+}$. For the K–V sample, the Al 2p peak at 73.7 eV corresponds to tetrahedral Al ($^{[4]}$Al) (orange), and higher binding energy peak of 74.4 eV corresponds to the octahedral Al ($^{[6]}$Al) (green) [Ref.[1,2]]. The Al 2p peak (both orange and green) is being shifted towards the higher binding energy side for Ca–V and Al–V samples. Mg 2s spectrum of K–V membranes shows Mg-O (purple) peak at 88.3 eV whereas the broad peak (light blue) can be ascribed to $Mg(OH)_2$ [Ref. [3]]. These peaks are also observed in the case of Ca–V and Al–V membranes.

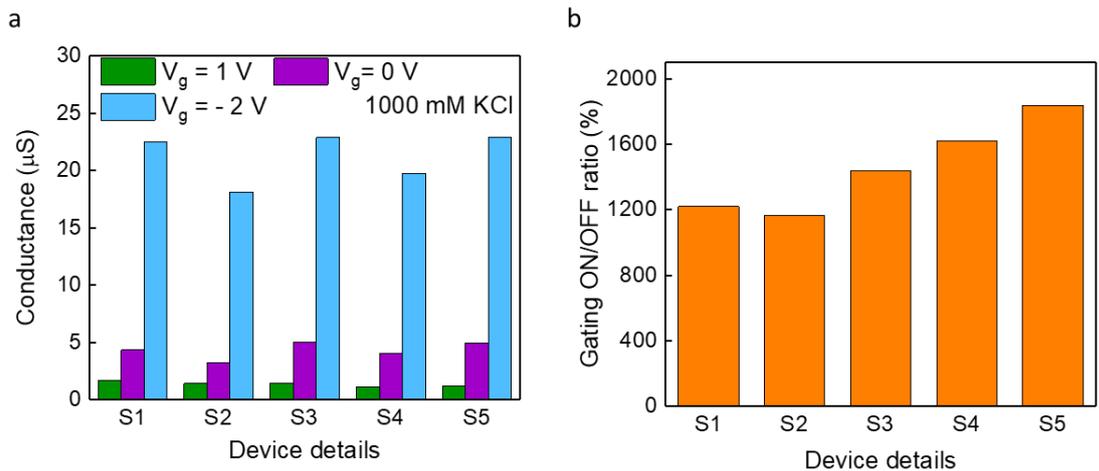

**Supplementary Figure 11. Repeatability and reproducibility of several voltage-gated K–V devices. a** Conductance of 5 different samples at $V_g$ = +1 V, 0 V, and -2 V for 1000 mM KCl. **b** The gating ON/OFF ratio estimated from Supplementary Fig. 11a, for the extreme gate voltages, is displayed for several samples. We observed that most of our devices were stable and showed a similar behavior.

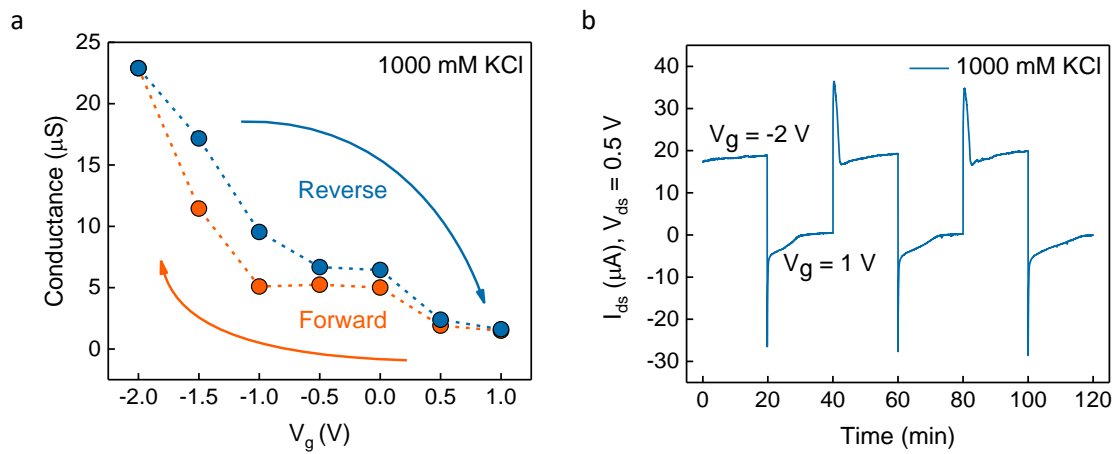

**Supplementary Figure 12. Reversible modulation of conductance for 1000 mM KCl in K–V laminates. a** Conductance change for 1000 mM KCl with gate voltages; the arrow represents the direction of the gate potential applied to the sample. The doted lines are a guide for the eye. This clearly shows that the device is stable, and at high concentration, we could tune the conductance efficiently. **b** Multicycle ionic current ($I_{ds}$) measurements at $V_g$ = -2 V and 1 V with $V_{ds}$ = 0.5 V shows the reversible modulation.

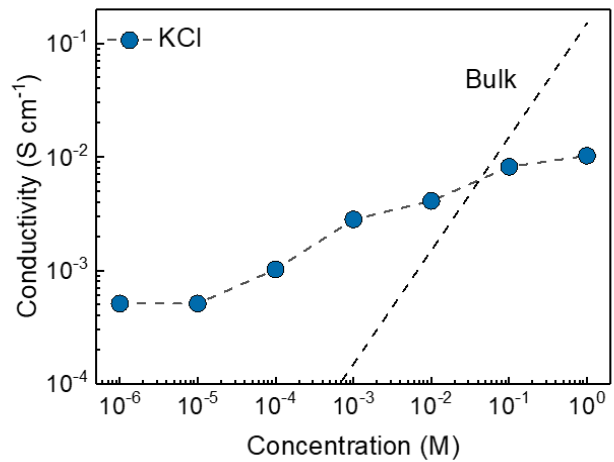

**Supplementary Figure 13. Concentration-dependent conductivity data of K–V laminates**. The measured conductance is plotted as a function of KCl concentration. The dotted line is a guide for the eye.

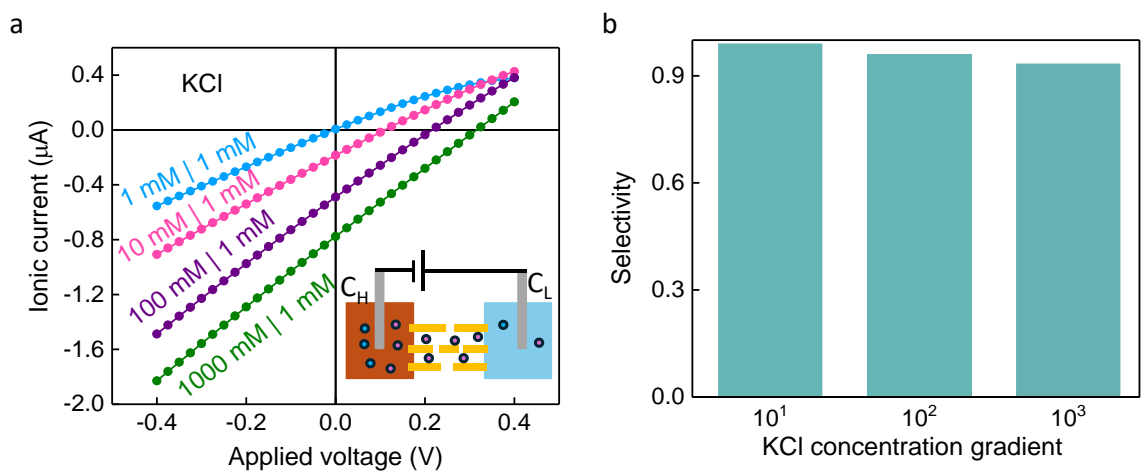

**Supplementary Figure 14. Diffusion studies of the K–V sample. a** The ionic current as a function of applied potential without subtracting the redox potential of the electrodes. Inset: schematic of the diffusion measurement setup. **b** Ion selectivity with different KCl concentration gradients.

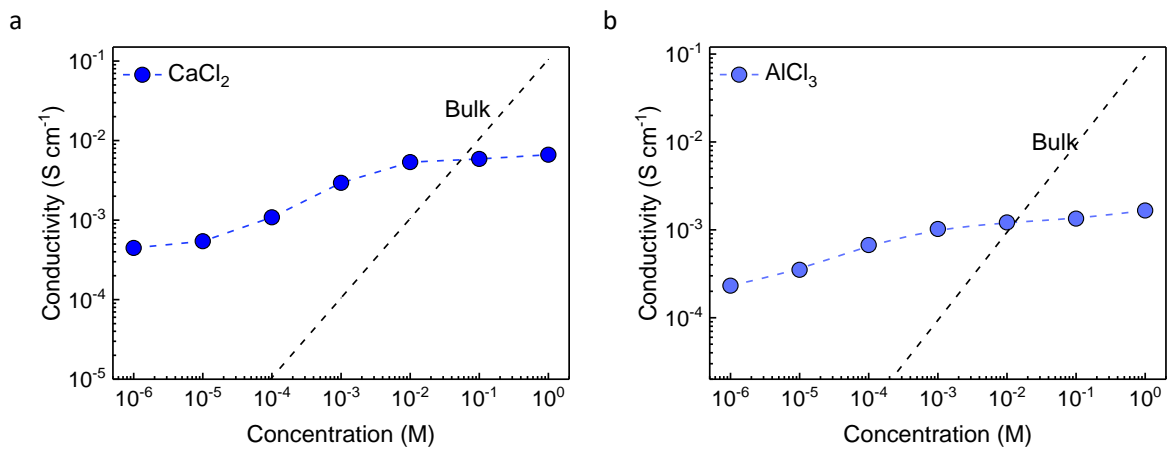

**Supplementary Figure 15. Conductivity variation with concentration. a** Ca–V laminate with CaCl$_2$ solution. **b** Al–V laminates with AlCl$_3$ solution. The dotted line is a guide for the eye.

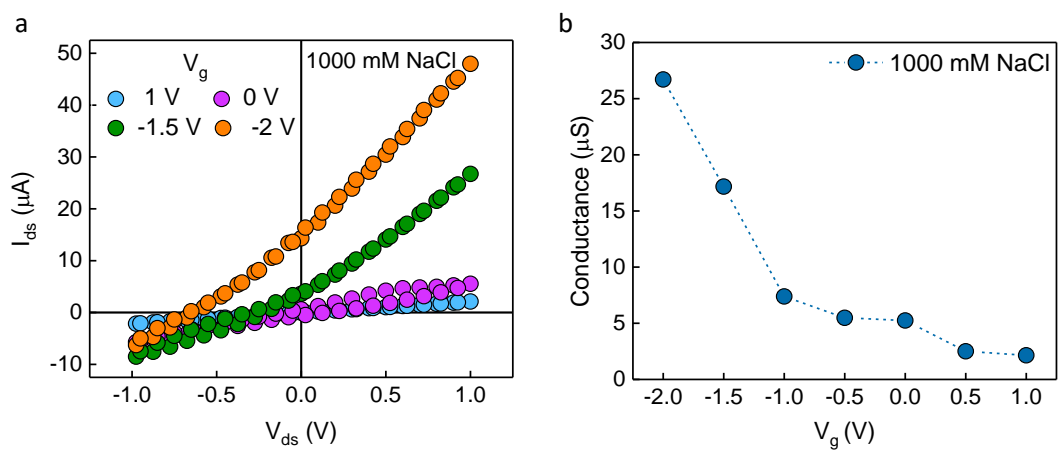

**Supplementary Figure 16**. **Voltage-gated ion transport through Na–V membranes with 1000 mM NaCl**. **a** Modulation of ionic current, I$_{ds}$ with different V$_g$ from -2 V to 1 V with a step size of 0.5 V. The NaCl solution concentration is 1000 mM. **b** Variation of conductance with gate voltage (V$_g$) at a concentration of 1000 mM NaCl. The gating ON/OFF ratio is found to be ~1100%, slightly smaller than KCl for the same concentration.

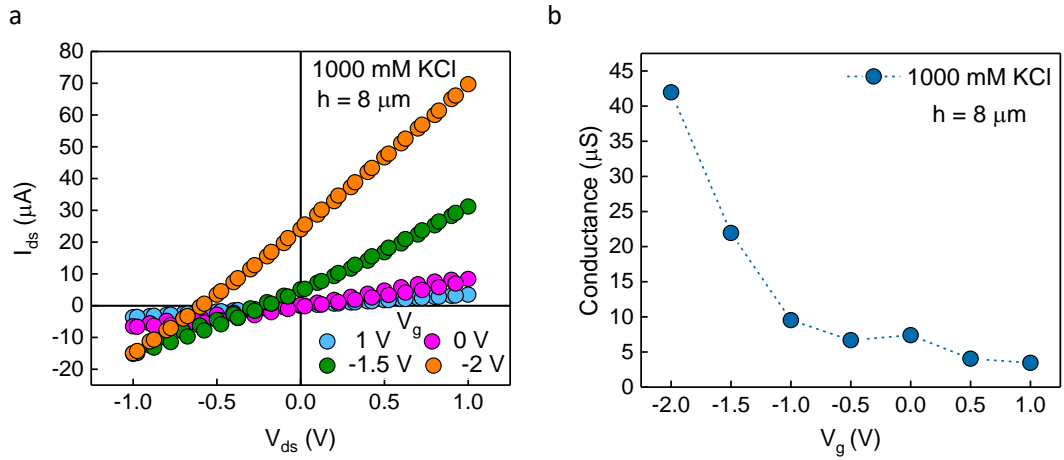

**Supplementary Figure 17**. **Voltage-gated modulation of KCl ion transport through K–V membrane of thickness, h = 8 μm at a concentration of 1000 mM**. **a** Modulation of ionic current, $I_{ds}$ with different $V_g$ at 1000 mM KCl. **b** Conductance variation with different $V_g$ at a concentration of 1000 mM KCl. The overall ionic conductance is higher in this case when compared to h = 3.5 μm. However, the gating ON/OFF ratio is found to be ∼1100%, slightly smaller than h = 3.5 μm.

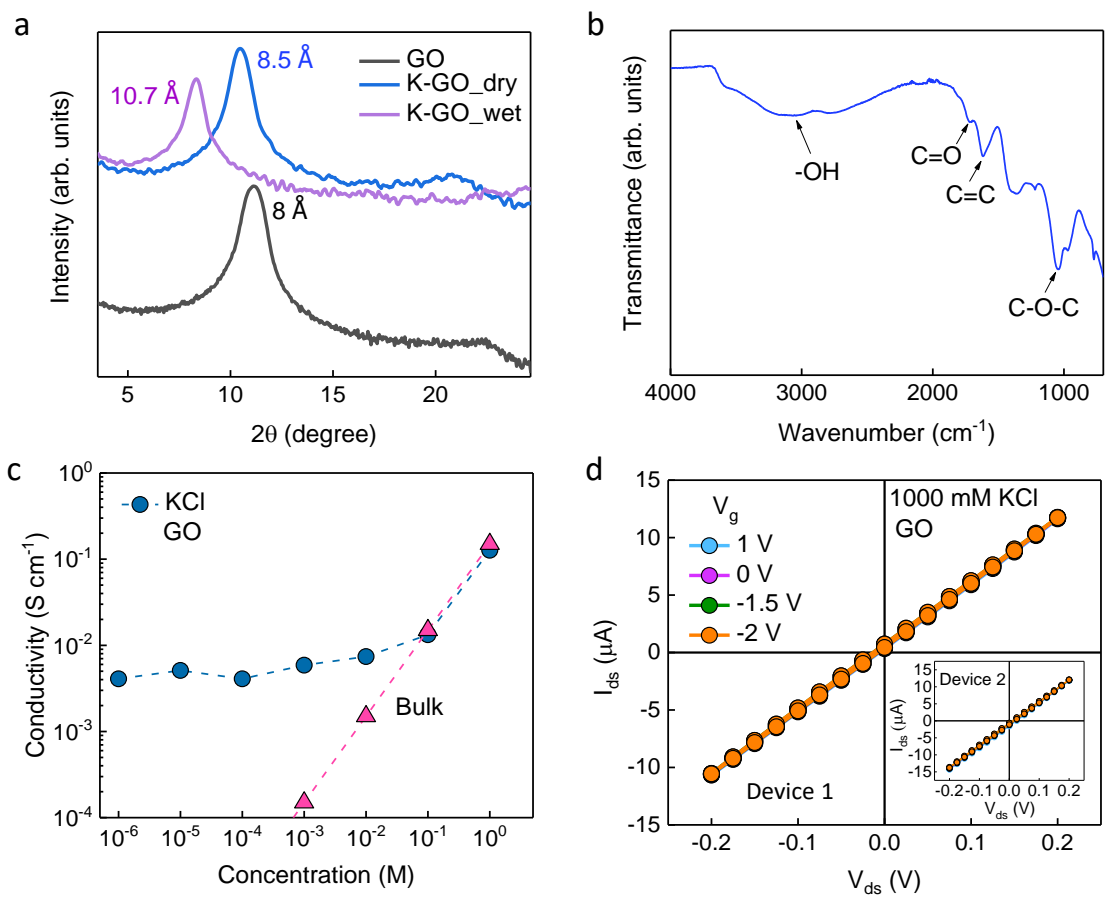

**Supplementary Figure 18. Characterization and gated ion transport studies through the GO sample a** XRD studies of GO, K-intercalated GO in wet and dry conditions. **b** FTIR studies of the GO sample show the characteristic peaks of -OH, C=C, C=O, and C-O-C. **c** Conductivity vs. concentration characteristics (represented as blue circles) of the GO sample with KCl salt solution. Bulk conductivity data (represented as pink triangles) is also provided for easy reference. **d** $V_{ds}$-$I_{ds}$ characteristics at 1000 mM KCl for a GO sample (device 1) with different $V_g$ values from -2 V to 1 V. Inset shows the data obtained from another device (device 2) with similar applied $V_g$ as device 1.

**Supplementary Note 1: Surface charge density calculation**

The surface charge density of the vermiculite layer is calculated from the measured zeta potential employing the Gouy-Chapman equation [ref.4];

$$\sigma = -\frac{\varepsilon_0\ \varepsilon_r \xi}{\lambda_d}\left(\frac{\sinh\left(\frac{F\xi}{2RT}\right)}{\frac{F\xi}{2RT}}\right) \quad \text{Supplementary Equation (1)}$$

where $\varepsilon_r$ is the dielectric constant, $\varepsilon_0$ is the permittivity of the free space, $\lambda_d$ is the Debye length, F is the Faraday constant, R is the gas constant, T is the temperature, and $\xi$ is the zeta potential. The surface charge density is found to be ~-4 mC/m² for K–V membranes, assuming $\varepsilon_r$ = 80 and C = 1 mM. The surface charge density for Ca–V and Al–V, are ~-1.6 mC/m² and ~-0.3 mC/m².

**Supplementary Note 2: Diffusion studies**

In order to determine the ion selectivity of the membranes, we have carried out drift-diffusion studies through our in-plane K–V devices. The same voltage-gated membrane device was placed in the middle separating two reservoirs. The reservoirs were filled with various salt solutions such that concentration gradients of 10, 100, and 1000 were established. For this, we fixed the concentration of one reservoir to 1 mM and varied the concentration of the second reservoir from 10 mM to 1000 mM. Two home-made Ag/AgCl electrodes were used to measure the *I-V*'s.

It is known that electrodes placed in different concentrations produce an additional voltage known as the redox potential $V_{redox}$, and can be calculated from the equation given below[5],

$$V_{redox} = \frac{k_B T}{e}\ln\left(\frac{\gamma_H C_H}{\gamma_L C_L}\right) \quad \text{Supplementary Equation (2)}$$

Where $\gamma_H$ and $\gamma_L$ are the mean activity co-efficient of ions at high and low salt concentrations, which depends on the type of salt used and their concentration. The actual membrane potential or diffusion potential, $V_{diff}$ in the case of Ag/AgCl electrodes can be obtained after deducting the redox potential from the measured potential ($V_{measured}$).

$$V_{diff} = V_{measured} - V_{redox} \quad \text{Supplementary Equation (3)}$$

The selectivity is calculated from the Nernst potential as follows,

$$V_{diff} = S\frac{RT}{zF}\ln\left(\frac{\gamma_H C_H}{\gamma_L C_L}\right) \quad \text{Supplementary Equation (4)}$$

Where R, T, and F are gas constant, temperature, and Faraday constant respectively, z is the ion valence. Here, S = $t_+$ - $t_-$, where $t_+$, and $t_-$ are the cationic and anionic transference numbers, respectively. Ion selectivity, S = 1, implies a perfectly cation-selective membrane, and S = 0 means the absence of any ion selectivity. We found that S = 0.96 for our K–V device from the above formula. This shows that K–V laminates are near ideal cationic membranes that only allow cations to transport even at high chloride concentrations of 1000 mM.

**Supplementary Note 3: Fabrication of graphene oxide membrane:**

**Materials**

Graphite, $KMnO_4$, $H_2O_2$ (35 wt%), HCl (34 wt%), potassium chloride (KCl) (> 99%), and PVDF (0.22 μm pore size) were purchased from Sigma Aldrich.

**Fabrication of GO laminates**

The synthesis of graphene oxide (GO) was carried out utilizing a modified Hummers' method. Initially, 1 g of graphite flakes was mixed with 35 mL of concentrated sulfuric acid ($H_2SO_4$) in a beaker. This mixture was subjected to stirring at 500 rpm using a magnetic stirrer, while ensuring that the temperature remained below 10°C. Subsequently, 3 g of potassium permanganate ($KMnO_4$) was incrementally added to the beaker, resulting in a color change of the solution to greenish. This mixture was stirred continuously for a duration of 6 hours. Following this period, 25 mL of deionized (DI) water was added dropwise to the slurry to terminate the oxidation process, after which an additional 50 mL of DI water was added. This resulted in a shift of the solution's color from greenish to brownish, an indication of the formation of graphite oxide. After allowing the mixture to settle for 15 minutes, 5 mL of hydrogen peroxide ($H_2O_2$) was added, which caused the color of the solution to return to a greenish tone. To purify the resulting slurry, a hydrochloric acid (HCl) solution at 10 wt.% was used, followed by centrifugation at 8000 rpm. This process was further continued with DI water until the pH of the solution reaches 7. The resultant GO powder was obtained by freeze-drying the slurry for 72 to 90 hours. A solution with a concentration of 1 mg/mL of GO was prepared using DI water as the solvent; this involved sonicating the mixture for 1 hour and subsequently centrifuging at 5000 rpm to isolate the supernatant for membrane preparation. Polyvinylidene fluoride (PVDF) with pore size of 0.22 μm was utilized as the supporting membrane during the fabrication process through vacuum filtration. After drying the GO membrane on PVDF, it was observed that the membrane easily peels off from the PVDF support. These membranes were further used for the characterization and ion transport studies.

**XRD and FTIR studies**:

The XRD of GO shows a d-spacing of 8 Å and a transport height of 4.6 Å after deducing the single-layer thickness of graphene (0.34 nm) (Supplementary Fig. 18a). The d-spacing of as-prepared GO is consistent with earlier reports[6]. The KCl (1 M) dipped GO membrane shows a d-spacing of 8.5 Å and 10.7 Å, respectively, in dry and wet conditions. Since our measured devices are in an aqueous solution, the transport height for the KCl-dipped GO membrane is 7.3 Å. We carried out an FTIR study of the GO sample, which shows characteristic peaks of -OH at 3088 cm$^{-1}$, C=C at 1617 cm$^{-1}$, C=O at 1720 cm$^{-1}$, and C-O-C at 1040 cm$^{-1}$, matching with literature values (Supplementary Fig. 18b)[7]. We performed ion transport studies through the as-fabricated membrane, and the conductivity vs. concentration characteristics show that above 100 mM concentration, the transport is bulk-like. As the transport height is 7.3 Å, at 1000 mM concentration, the Debye layer does not overlap and gives a bulk conductance value (Supplementary Fig. 18c). Then we checked the effect of gating at 1000 mM KCl concentration, and the I-V results are displayed with different $V_g$ values; the slope did not change, suggesting that the conductance is not influenced by a gate (Supplementary Fig. 18d).

**Supplementary Table 1. Summary of gating ON/OFF ratio for different gated nanofluidic systems.** Most of the studies were performed at dilute salt concentrations, and the large Debye screening length at these concentrations ensures that there is an electrostatic effect.

| Material | Concentration | Transport height | Gate voltage ($V_g$) | Gating ON/OFF ratio (%) | Reference |
|---|---|---|---|---|---|
| GO | 50 mM KCl | 1.66 nm | -0.5 V to 0 V | ~300 - 400 | *Nature Nanotech* 13, 685 (2018) |
| MXene | 10 mM KCl | 0.7 nm | 1 V to -1 V | <1000 | *ACS Nano* 13, 11793 (2019) |
| Mesoporous silica nanotube aligned film | [$H^+$] = 0.1 mM | <8 nm | -1 V to 1 V | ~200 | *Nature Mater* 7, 303 (2008) |
| rGO | 0.2 M KCl | ~0.3 nm | 0 V to -1.2 V | Off to On | Science 372, 501 (2021) |
| Silica nano channel | 1 mM KCl | 35 nm | -20 V to 20 V | ~130 | Nano Lett. 5, 943 (2005) |
| Vermiculite | 1000 mM KCl | ~0.3 nm | -2 V to 1 V | 1400 | This work |

**Supplementary Table 2. Activity coefficients and ion selectivity of K–V membranes in KCl solutions.** The parameters ion selectivity, $S$ and activity coefficient, $\gamma$ with KCl solution is provided for K–V membranes at different concentration gradients. The redox potential is estimated from the Nernst equation and provided in the table at different concentration gradients. These measurements were performed at a temperature of 298 K.

| Concentration gradient (KCl) | $V_{measured}$ (mV) | $V_{redox}$ (mV) | $V_{diff}$ (mV) | $\frac{\gamma_H}{\gamma_L}$ (Ref. [8]) | S |
|---|---|---|---|---|---|
| 10 | 114.20 | 57.37 | 56.83 | 0.933 | 0.9906 |
| 100 | 220.40 | 112.40 | 108 | 0.795 | 0.9608 |
| 1000 | 319.90 | 165.37 | 154.53 | 0.625 | 0.9345 |